\documentclass[showkeys,longbibliography, showpacs,nopreprintnumbers,prd,superscriptaddress,nofootinbib,floatfix]{revtex4-1}% for review and submission
\usepackage{graphicx}  % needed for figures
\usepackage{dcolumn}   % needed for some tables
\usepackage{bm}        % for math
\usepackage{physics}
\usepackage{amssymb}   % for math
\usepackage{amsmath,amsthm,latexsym,amssymb,amsfonts,epsfig}
\usepackage{hyperref}
\usepackage[capitalize]{cleveref}
\usepackage{float}
\usepackage{adjustbox}
\usepackage{multirow}
\usepackage{comment}
\usepackage{tabularx}
\usepackage[usenames,dvipsnames]{xcolor}
\newcommand{\be}{\begin{equation}}
\newcommand{\ee}{\end{equation}}
\newcommand{\bea}{\begin{eqnarray}}
\newcommand{\eea}{\end{eqnarray}}
\begin{document}

% The following information is for internal review, please remove them for submission
\widetext
%\leftline{Version xx as of \today}
%\leftline{Primary authors: Micha{\l} Artymowski, Ido Ben-Dayan, Utkarsh Kumar}
%\leftline{To be submitted to PRL}
%\leftline{Comment to {\tt d0-run2eb-nnn@fnal.gov} by xxx, yyy}
%\centerline{\em D\O\ INTERNAL DOCUMENT -- NOT FOR PUBLIC DISTRIBUTION}

% the following line is for submission, including submission to the arXiv!!
%\hspace{5.2in} \mbox{Fermilab-Pub-04/xxx-E}

\title{Probing The Early Universe Cosmology With NANOGrav: Possibilities and Limitations}
%\input author_list.tex       % D0 authors (remove the first 3 lines
                             % of this file prior to submission, they
                             % contain a time stamp for the authorlist)
                             % (includes institutions and visitors)
\author{Ido Ben-Dayan}
\affiliation{Physics Department, Ariel University, Ariel 40700, Israel}

\author{Utkarsh Kumar}
\affiliation{Physics Department, Ariel University, Ariel 40700, Israel}

\author{Udaykrishna Thattarampilly}
\affiliation{Physics Department, Ariel University, Ariel 40700, Israel}
\author{ Amresh Verma }
\affiliation{Physics Department, Ariel University, Ariel 40700, Israel}
\date{\today}

\begin{abstract}
A stochastic gravitational wave background is a prediction of a number of astrophysical and cosmological phenomena including early universe cosmology. 
Recently, the NANOGrav Collaboration reported conclusive evidence for a stochastic gravitational-wave background. We analyze the NANOGrav signal assuming it is of primordial origin including the reheating phase. %Analysis of the NANOGrav signal has shown that if the signal is primordial, then the power spectrum of primordial gravitational waves is blue-tilted. However, this analysis does not take into account of the reheating phase and a number of other factors like running of the spectrum. The era of reheating is the poorly understood epoch in the expansion history of the Universe at the end of inflation. We explore the possibility of constraining the era of reheating assuming NANOGrav signal is primordial. 
We use the latest measurements from NANOGrav to constrain the Universe's reheating equation of state $w_{re}$, the reheating temperature $T_{re}$, the tensor to scalar ratio $r$, and the tensor tilt $n_t$. Assuming the constant equation of state $w_{re}$ responsible for reheating phase, we find preference for instant reheating, $w_{re} = 0.36^{+0.15}_{-0.28}$, and a very blue tilt  $n_t = 1.94^{+0.43}_{-0.88}$. 
We find a degeneracy between the tensor to scalar ratio $r$ and $T_{re}$ and suggest ways to break this degeneracy. In all cases where the reheating temperature is constrained, it is constrained to be very low with $T_{re}\leq 10^5 \, GeV$. We further find that a scale-invariant spectrum as suggested by inflation implies a stiff equation of state $w_{re}=19/3$.
If extrapolated, the blue-tilted primordial spectrum that agrees with the NANOGrav signal at corresponding frequencies is incompatible with the LIGO bound. This incompatibility is another challenge for connecting NANOGrav with the primordial spectrum. We discuss a number of ways to circumvent this issue. We split the spectrum into a sum of astrophysical and primordial spectra and constrain the astrophysical and primordial components using NANOGrav data and the LIGO bound. In another attempt, we use the same data and constrain the running of the spectrum. Any of these or a combination of such methods can be used to reconcile the NANOGrav data and the LIGO bound with the primordial power spectrum. 
\end{abstract}

\maketitle

\section{Introduction} \label{sec:intro}
A stochastic gravitational wave background (SGWB) is a robust prediction of a number of well-motivated theories in cosmology including theories of the early Universe \cite{Caprini:2018mtu,Renzini:2022alw}. These theories generically predict SGWB that spans a large interval of frequencies. Pulsar timing arrays (PTAs) are a well-known detection channel for nHz gravitational waves (GW).  The arrays operate by detecting the spatial correlations in the pulse arrival time for pulsars due to the perturbations in space-time induced by GWs \cite{Hobbs_2017}. In 2020, the NANOGrav Collaboration published an analysis of 12.5 yrs of pulsar timing data \cite{NANOGrav:2020bcs} reporting strong evidence for a stochastic common-spectrum process. This signal was later confirmed by a number of collaborations, Parkes PTA (PPTA) \cite{Goncharov:2021oub}, European PTA(EPTA) \cite{Chen:2021rqp}, as well as the
combined International PTA (IPTA) \cite{Antoniadis:2022pcn}. Recently in 2023 various PTA Experiments including NANOGrav, EPTA, PPTA, an CPTA (Chinese PTA) confirmed the presence of excess red common-spectrum signals with an amplitude of order $10^{-15}$ at a reference frequency range around $1 yr^{-1}$ \cite{NANOGrav:2023gor,Antoniadis:2023rey,Reardon:2023gzh,Xu:2023wog,NANOGrav:2023hde,agazie2023nanograv,NANOGrav:2023tcn,NANOGrav:2023hvm,NANOGrav:2023ctt,johnson2023nanograv,NANOGrav:2023pdq,Antoniadis:2023puu,Antoniadis:2023aac,Antoniadis:2023xlr,EuropeanPulsarTimingArray:2023qbc,Reardon:2023zen,Zic:2023gta}. More importantly, all recent analyses point to a GW origin of the signal.
Having nearly confirmed that the measurements have a GW origin, the exciting possibility that this background signal is being sourced by fluctuations in the early Universe has been discussed in the literature \cite{10.1093/mnrasl/slaa203,Vagnozzi:2023lwo,Datta:2023vbs,Lazarides:2023ksx,Zhao:2023joc,Das:2023nmm,Gorji:2023sil,Ahmadvand:2023lpp,Zhang:2023nrs,Bousder:2023ida,ValbusaDallArmi:2023nqn,Cui:2023dlo,Basilakos:2023xof,Balaji:2023ehk,Gelmini:2023kvo,Yamada:2023thl,Babichev:2023pbf,Buchmuller:2023aus,You:2023rmn,Salvio:2023ynn,Gouttenoire:2023bqy,Geller:2023shn,Du:2023qvj,Servant:2023mwt,Wu:2023hsa,Li:2023tdx,Liu:2023pau,Liu:2023ymk,Oikonomou:2023qfz,Figueroa:2023zhu,Unal:2023srk,Niu:2023bsr,Cai:2023dls}.

Models of the early Universe such as inflation were introduced in order to solve problems with standard cosmology such as the horizon and flatness problem. Tensor and scalar fluctuations in the early Universe grow, stretching outside the causal horizon and re-entering the horizon at a much later time. These fluctuations suggest that the origin of the structure is quantum and predict the CMB spectrum with great accuracy. %In spite of a few potential foundational problems pointed out recently (\cite{Ijjas:2013vea,Ijjas:2014nta,Obied:2018sgi,Agrawal:2018own,Bedroya:2019tba,Palti:2019pca}) inflation has remained the most consistent theory of the early Universe. 
Tensor perturbations produced in the early universe are robust signatures of different early universe paradigms and are rather insensitive to the details of the different models \cite{Ben-Dayan:2016iks,Ben-Dayan:2018ksd}. These paradigms are important contenders for the source of SGWB. The detection and characterization of the SGWB result can serve to distinguish various models and paradigms of the early universe.

In \cite{Vagnozzi:2023lwo} the author explores the possibility of inflationary origins of the nHz GW signal and concludes that a standard slow-roll inflationary origin is less tenable. He concludes that for the nHz GW signal to have primordial origins, the power spectrum has to be blue-tilted with a spectral index of $n_t=1.8\pm 0.3 $. However, this analysis assumes instant reheating of the Universe after the inflationary phase. The effects of reheating on the GW spectrum of inflation have been discussed in literature over the years \cite{Martin:2014nya,Dai:2014jja,Lozanov:2019jxc}. It has been observed that the spectral index and amplitude of the GW signal are affected by the duration of the reheating era and also by the equation of state $w_{re}$, during the reheating phase \cite{Boyle:2007zx}. In this article, we examine the possibility of a primordial origin of the nHz GW signal in the presence of a non-trivial reheating phase, while remaining agnostic about the early universe paradigm that had been realized in Nature. 

We discuss the tenability of a primordial origin of the nHz signal in this renewed context. We perform a likelihood analysis of the data based on compressed likelihoods from \cite{Vagnozzi:2023lwo}. Slow-roll inflation, where $n_t\simeq 0$ will require a stiff reheating phase of $w_{re}\simeq 19/3$. A blue tilted spectrum with $n_t\sim 1.77$ is obtained for instant reheating in accordance with what we expect from previous works. Finally, open reheating equation of state $w_{re}$ and tensor tilt $n_t$ result in $n_t=1.94^{+0.43}_{-0.88}$, and $w_{re}=0.36^{+0.15}_{-0.28}$. In all cases, the reheating temperature is expected to be ``low", $T_{re}\leq 10^5 \, GeV$. Thus,  within the primordial interpretation of the signal, it seems instantaneous reheating and a primordial blue-tilted tensor spectrum are favored. This disfavors the standard canonical single field slow-roll inflationary scenarios and favors other paradigms like a bounce \cite{Battefeld:2014uga,Khoury:2001wf}, non-standard inflation (e.g. Galileon fields, beyond slow roll, etc.) \cite{Kobayashi:2010cm,Tahara:2020fmn,Piao:2004tq,Gruzinov:2004ty,Satoh:2008ck,Mishima:2019vlh,Cai:2014uka,Gong:2014qga,Chen:2013aj,Lin:2013sja,Gumrukcuoglu:2012wt} or GW spectrum generated by sourced fluctuations \cite{Barnaby:2011qe,Ben-Dayan:2016iks,Ben-Dayan:2018ksd,Wang:2014kqa,Caprini:2014mja,Mukohyama:2014gba,Bastero-Gil:2014oga,Cook:2011hg,Carney:2012pk,Senatore:2011sp,Matarrese:1997ay,Biagetti:2013kwa}. We briefly discuss the possibility of a bounce being the origin of the GW signal in the future directions section.

Upon assuming that the GW background we have observed is primordial, we find that the blue-tilted spectrum with a large spectral index will violate the bounds on the GW spectrum set by LIGO. Few mechanisms have been proposed in the literature such as a break in power spectrum that could explain this phenomenon. However, most of them are fine-tuned and involve a further complication of the model. %A much more natural approach is attributing only part of the signal as a primordial one, the rest being attributed to astrophysics such as incoherent black hole mergers \cite{NANOGrav:2023gor}.
We propose two purely phenomenological analyses that can alleviate this issue.  One method is assuming that the spectrum can be resolved into astrophysical and primordial components and another method is assuming a significant value for the running of the spectrum. The splitting into an astrophysical part and a primordial part is very natural, and we estimate the ratio between the primordial signal and the total signal. If the astrophysical and primordial components share the same power law, then the primordial signal is at most $14.09\%$ of the signal. Inserting the LIGO result reduces it further to a minuscule part well within the reported error bars. If we treat the astrophysical component as some fixed amplitude we find that the primordial signal is between 34 \% to 55 \% of the PTA signal. The second method is that of running. The idea of running is also natural as the constant spectral tilt is just an approximation. We find that the NANOGrav signal can be primordial and in accord with the LIGO bound with the running parameter, $\alpha_t\sim -0.08$. 
This paper is a rough report of our analysis, and we show that it is too early to conclusively forgo the possibility of the NANOGrav signal being primordial, though such a scenario is not favored. 

This article is organized as follows. In section \ref{sec:gwaves} we review the gravitational wave spectrum in the presence of a nontrivial reheating phase. In section (\ref{sec: pulsar}) we discuss the spectrum of PTA signals and connect it to the primordial spectrum. Section (\ref{sec: analysis}) deals with the details of our analysis and results. The incompatibility of NANOGrav signal with LIGO bounds and methods of resolution for this potential problem is discussed in section (\ref{sec:nli}). Finally, We conclude our findings and discuss the future directions in section (\ref{sec: conclusions}).

\section{Gravitational Waves in Early Universe and Reheating} \label{sec:gwaves}

Primordial GW is one of the most exciting possible explanations for SGWB. In this section, we briefly derive the gravitational waves spectrum generated in the early universe. We start our discussion by considering the line element of a perturbed FLRW metric in the synchronous gauge as:
\begin{equation}
    ds^2 = a^{2} (\eta)  \,\left[ d \eta^2 - \left(\delta_{ij} + h_{ij} \right) \, dx^{i}\,dx^{j} \right]
\end{equation}
Here $a$ and $\eta$ are the scale factor and the conformal time respectively, and $h_{ij}$ is the traceless-transverse symmetric $3 \times 3$ matrix describing the GW. The waves have two polarizations $+, \times$. Each of which is governed by the following equation in Fourier space:
\begin{eqnarray}
    h_k '' + 2 \mathcal{H}\,h_k' + k^2\,h_k = 0 \label{eq: gw}
\end{eqnarray} 
where $'$ denotes a derivative with respect to conformal time, $\mathcal{H}$ is the conformal Hubble parameter and $k$ stands for the wave number. For a given background one can solve eq.(\ref{eq: gw}) and calculate the primordial tensor power spectrum given by
\begin{eqnarray}
    \mathcal{P}_{T}^{\text{prim}} &=& 8 \frac{k^{3}}{2\,\pi^{2}} |h_k|^{2} . 
\end{eqnarray}
For effective model selection, we work with the following parametrization:
\begin{eqnarray}
    \mathcal{P}_{T}^{\text{prim}} (k) = r\, A_{s} \,\left(\frac{k}{k_{*}}\right)^{n_t}.
\end{eqnarray}
 Here $r$ and $A_s$ are the tensor to scalar ratio and amplitude of primordial scalar power spectrum evaluated at suitable pivot scale $k_*$ that varies from one experiment to another. We fix this pivot scale $ k_* = 0.05 Mpc^{-1} $  in accordance with Cosmic Microwave Background measurements from Planck \cite{Planck:2018vyg}. 
Gravitational Waves experiments such as laser-interferometer (LI) and PTA measurements report their results in terms of gravitational-wave energy spectrum measured today ($\tau = \tau_0$) as 
\begin{eqnarray}
    \Omega_{\text{GW}}^{\text{prim}} (f) &=& \frac{1}{\rho_{0}^{crit}}\,\frac{d \rho_{0}^{\text{GW}}}{d \ln f},
\end{eqnarray}
 where $f = \frac{9.72 \times  10^{-15}}{2 \,\pi \, a_0}  \frac{k}{\text{Mpc}^{-1}} \text{Hz}$ is the present-day physical frequency of a GW associated with the comoving wavenumber $k$. The present-day GW energy spectrum $\Omega_{GW}^{\text{prim}}(f)$ is then related to the primordial tensor power spectrum  $ \mathcal{P}_{T}^{\text{prim}} (f)$ via
 \begin{eqnarray}
     \Omega_{\text{GW}}^{\text{prim}} (f) &=& \frac{1}{12} \, \left[ \frac{2 \pi f}{H_0}\right]^{2} \, T_{h} (f) \, \mathcal{P}_{T}^{\text{prim}} (f),
 \end{eqnarray} 
where 
$T_{h} (f)$ is  the ``tensor transfer function" derived in \cite{Boyle:2007zx}. The tensor transfer function is expressed in the following form
 \begin{eqnarray}
     T_{h} (f) & = & \frac{C_2 (f)\, C_3 (f) }{2 \, \left(1 + z_{re}\right)^{2}} \, \left[ \frac{\Tilde{\gamma}^{-1/2}\, 2\,\pi\,f}{\left(1 + z_{re} \right)\, H_0}\right]^{-4 / (1 + 3\,w_{re} (f))} \,  \label{eq: tensortransfer}
 \end{eqnarray}
 Using the tensor transfer function $T_{h} (f)$ mentioned in eq. \eqref{eq: tensortransfer}, the present-day GW energy spectrum is written as 
 \be
          \Omega_{\text{GW}}^{\text{prim}}(f) = \frac{r \, A_s\, C_2(f)\, C_3 (f)\, \Tilde{\gamma} }{24} \, \left(\frac{\Tilde{\gamma}^{-1/2}\,  2\,\pi\,f}{(1 + z_{re})\,  H_0}\right)^{2\,\frac{3 w_{re}(f) -1}{3 w_{re}(f) +1}}  \, \left(\frac{a_0\,H_0\,2\,\pi\,f}{k_{*}\,H_0}\right)^{n_t(f)}, \label{eq: GWInfl}
 \ee 
where the factors \textcolor{Blue}{$\Tilde{\gamma}$}, $C_2 (f)$ and $C_3 (f)$ are defined as 
\begin{eqnarray}
    \Tilde{\gamma} = \frac{\Omega_{m0}}{1 + z_{eq}} \frac{g_{*}(z_{re})}{g_{*}(z_{eq})}  \frac{g_{*s}^{4/3}(z_{eq})}{g_{*s}^{4/3}(z_{re})}
\end{eqnarray}
\begin{eqnarray}
       C_2 (f) = \frac{\Gamma^{2} \left( \frac{5 + 3w_{re}(f)}{2(1 + 3w_{re}(f))}\right)}{\pi } \, \left[ 1 + 2 w_{re}(f) \right]^{\frac{4}{1 + 3\, w_{re}(f)}} \,, \cr
    C_3(f) = \left[- \frac{10}{7}\frac{(98\,\Omega_{fs}^{3} - 589\,
    \Omega_{fs}^{2} + 9380\,\Omega_{fs}- 55125)}
  {(15 + 4\,\Omega_{fs})(50 + 4\,\Omega_{fs})(105 + 4\,\Omega_{fs}^{})}\right]^{2}
\end{eqnarray}
 with $\Omega_{m0}$ denoting the relative present-day matter energy density, $\Omega_{fs} \equiv \frac{\rho^{fs}}{\rho^{crit}} $ is the fraction of critical density which is free-streaming relativistically (e.g., neutrinos) at redshift $z_{re}$, and $g_{*}(z), \, g_{*s}(z)$ are the number of effective relativistic degrees of freedom at
the redshift $z$, as measured by the energy density $\rho(z)$ or the
entropy density $s(z)$, respectively \cite{Boyle:2007zx}. %$\rho_{m0}$ to the present-day
%critical density $\rho_0^{crit}$ 
In the above expressions, $w_{re}$ and $z_{re}$ are the reheating equation of state (eos) and reheating redshift, i.e., $z$ at which the universe becomes radiation dominated. Also in the expression of $\Omega_{GW}$, we can replace parameter $1+z_{re}$ with $T_{re}$ in GeV, i.e., the temperature at the end of the reheating era using the following relation:
\begin{eqnarray}
    \frac{1+z_{re}}{1+z_{eq}} &=& \frac{T_{re}}{T_{eq}} \nonumber \\
    1+ z_{re} & \approx & \frac{3400 \times 10^{11}}{80} \frac{T_{re}}{GeV}.
\end{eqnarray}
where we have used the numerical value of $z_{eq}$ and $T_{eq}$ from Planck 2018 data \cite{Planck:2018vyg}.

\section{Pulsar timing arrays and connection to Early Universe } 
\label{sec: pulsar}
This section reviews the SGWB spectral energy density associated to modes relevant to PTA. The present-day PTA GW spectral energy density is expressed as:
\begin{eqnarray}
    \Omega_{GW}^{\text{PTA}} (f) &=& \frac{2 \pi^2}{3 H_0^2} \, f^2 \, h_{c}^{2}(f) ,
\end{eqnarray}
where $h_{c}(f)$ is the power spectrum of GW at PTA scales, measured at the reference frequency, $f_{yr} = 1 yr^{-1} \approx 3.17 \times 10^{-8} Hz$. $h_{c}(f)$ is parameterized in the following way:
\begin{eqnarray}
    h_c (f) &=& A \, \left(\frac{f}{f_{yr}}\right)^{\beta}\,,
\end{eqnarray}
where $A$ and $\beta$ are the amplitude and spectral index associated with the PTA signal. The spectral index $\beta $ is related to  timing-residual cross-power spectral density index $\gamma$  via $(3 - \gamma) / 2$. The 15-yr NANOGrav data measured the characteristic strain in the frequency range $f \in [2.0 \times 10^{-9}, 6 \times 10^{-8}]$ and this data is fitted as power law using the following corresponding GW energy density $\Omega_{GW}^{\text{PTA}}$ 
\begin{eqnarray}
    \Omega_{GW}^{\text{PTA}}(f) &=& A^{2} \frac{2\pi^{2}}{3H_0^2} \, \frac{f^{5 -\gamma}}{f_{yr}^{\gamma -3}} \,.
    \label{eq: GWPTA}
\end{eqnarray}

PTA data from NANOGrav gives constraints on $A$ and $\gamma$ as a joint posterior distribution. The collaboration inferred the values of $A = 6.4 ^{+4.2}_{-2.7} \times 10^{-15}$ and $\gamma = 3.2 \pm 0.6$ at 2 $\sigma$ CL. Let us connect these quantities to early universe parameters. Equating the eq. \eqref{eq: GWInfl} with eq. \eqref{eq: GWPTA} gives the following relation of $\gamma$ with $n_t$ and $w_{re}$ 
\begin{eqnarray}
    \gamma &=& 5 - n_t + 2 \, \alpha, \label{eq:gammaPTA}
\end{eqnarray}
where we have defined $\alpha = \left( \frac{1 - 3\,w_{re}}{1 + 3\,w_{re}} \right)$.
Similarly, the amplitude of the PTA signal is governed by the following equation
\be
    A = \left( \frac{A_s \,C_2 \, C_3 \, H_0^2}{16\,\pi^2} \right)^{1/2} \,  \tilde \gamma^{\left(1 + \alpha\right) / 2} \,  \left(2\,\pi\right)^{ \left(n_t - 2\,\alpha\right) / 2}\,  \left(\frac{a_0}{k_{*}}\right)^{n_{t} /2}\,\left(\frac{T_{\text{eq}}}{H_0 \left(1 + z_{\text{eq}}\right) T_{\text{re}}}\right)^{-\alpha} \, \sqrt{r}\, yr^{\left(1 + \alpha - n_t / 2\right)} \label{eq:APTA}
\ee
 Note that we have dropped the frequency dependence on $n_t$ and $w_{re}$ for simplicity. We will discuss the implications of frequency dependence on reheating eos later. The functional form of $A$ has dependence on $n_t$, $r$ and $T_{re}$. We insert the best-fit values for the cosmological parameters, i.e. $\mathcal{P}_T (k_{*}, \tau_{i}) \equiv A_s = 2.1 \times 10^{-9}$ at $k_{*} = 0.05 \text{Mpc}^{-1}$, $k_{eq} \approx 0.01\, \text{Mpc}^{-1} $, $z_{eq} = 3400$, $T_{eq} \approx 1.25 \times 10^{29}\, \text{Mpc}^{-1} $, $\tilde \gamma \approx 8.03 \times 10^{-5}$ from the Planck 2018 results \cite{Planck:2018vyg}. It is worth mentioning that our discussion about reheating neglects transient periods from inflation to the new eos $w_{re}$ and if $w_{re}\neq 1/3$ then the later transition to radiation domination. We assume these periods were very short with practically no effect on observables. Thus, $w_{re}=1/3$ means ``instantaneous reheating" while $w_{re}\neq 1/3$ means a finite reheating period which ends at $T_{re}$, where the Universe becomes radiation dominated.
Moreover, reheating parameters impact the radiation energy density of the early Universe from GWs, and result in an additional contribution. This contribution is quantified by $\Delta N_{eff}$ which affects predominantly at the scale of Big Bang Nucleosynthesis (BBN) and recombination. The SGWB contribution to $\Delta N_{eff}$  is characterized by \cite{Kuroyanagi:2014nba,Ben-Dayan:2019gll,Vagnozzi:2022qmc,Giare:2022wxq}
\begin{eqnarray}
    \Delta N_{eff}^{GW} \approx 1.8 \times 10^{5}\, \int_{f_{min}}^{f_{max}} df \frac{\Omega_{GW} (f) \, h^2}{f}\label{eq:Neff}
\end{eqnarray} 
where the values of $f_{min}$ and $f_{max}$ depend on the epoch of interest and the maximum temperature  reached in the Big Bang era. We will discuss the implications of reheating models on integral (\ref{eq:Neff}) in the next section.
\section{Data Analysis and Discussion} \label{sec:Discussion}
\label{sec: analysis}
We use the latest NANOGrav 15-year dataset to constrain the early universe parameters. In order to extract the parameter constraints using the NANOGrav observations, we follow the analysis done in \cite{Vagnozzi:2023lwo}. In \cite{Vagnozzi:2023lwo}, the author translated the inferred NANOGrav 15-year constraints on $A$ and $\gamma$ into constraints on inflationary parameters ($r$,$ n_t$) using a grid scan. We further generalize the analysis by taking into account the effect of the era of reheating. This is realized using a constant eos $w_{re}$ and reheating temperature, $T_{re}$. The data can be effectively approximated by using a Gaussian prior on $\log_{10} A$ and $\gamma$, with the following mean vector $\mu_{15}$ and covariance matrix $\Sigma_{15} $ as
\begin{eqnarray*}
    \mu_{15} &\approx & \begin{pmatrix}
        -14.20, 3.20
    \end{pmatrix} \\
    \Sigma_{15} & \approx & \begin{pmatrix}
        0.127 & -0.045 \\
        -0.045 & 0.021 
    \end{pmatrix}
\end{eqnarray*}
Using the mean vector $\mu_{15}$ and covariance matrix $ \Sigma_{15}$, we define the log-likelihood function for NANOGrav 15-year dataset as follows:
\begin{eqnarray} \label{eq: likelihood}
    \ln \mathcal{L} (\theta) = - \frac{1}{2} \, \left(x (\theta) - \mu_{15}\right)^{T}\, \Sigma_{15}^{-1}\, \left(x (\theta) - \mu_{15}\right).
\end{eqnarray} 

Here $ x(\theta)$ is the vector of parameters $x(\theta) \in (\log_{10} A (\theta), \gamma (\theta))$, where $\theta $ stands for the vector of inflationary and reheating parameters. We add the NANOGrav log-likelihood given in eq.(\ref{eq: likelihood}) to \texttt{Cobaya} \cite{Torrado:2020dgo} and generate the MCMC chains with convergence criterion $R - 1 < 0.001$. We work with $\log_{10} r$ and $\log_{10} T_{re}$ instead of $r$ and $T_{re}$ due to the statistical reasons. There exists a degeneracy between the $\log_{10} r$ and $\log_{10} T_{re}$ as both of them contribute to the amplitude of gravitational waves. This degeneracy can be broken in two ways. First, we can define a new parameter using the combination of $\log_{10} r$ and $\log_{10} T_{re}$ as $x_{re} = \log_{10} \left( r \, T_{re}^{2 \frac{1 + 3 w_{re}}{1 - 3 w_{re}}}\right)$. The second way is to fix the value of reheating temperature ($T_{re}$) / reheating eos ($w_{re}$). We explore both possibilities and illustrate our findings. Moving to the prior choices, we choose a flat prior on $ \log_{10} r \in [-25, -1.44]$, $ n_t \in [0,10]$, $w_{re} \in [-1/3, 1]$ and $\log_{10} T_{re} / \text{GeV} \in [-5, 15]$. We imposed the higher bound on $\log_{10} r $ such that it respects the current $95 \%$ upper limit reported from the joint analysis of Planck, WMAP, BICEP2, and BICEP3 data \cite{BICEP:2021xfz}.
    \begin{table}[H]
    \centering
68 $\%$  Constraints from NANOGrav 15-year data for reheating scenarios
   \vspace{1 em}  \\
{
\begin{tabular}{|c|c|c|c|c|}

\hline
\hline

    Parameter         & $w_{re}$ free &  $w_{re} = 0$ & $w_{re} = 1/3$ &  $ w_{re} = 1$ \\
\hline
\hline
$ n_t $        & $ 1.94^{+0.43}_{-0.88} $ & $3.72^{+0.23}_{-0.071} $& $ 1.77^{+0.18}_{-0.12}$ & $ 0.82^{+0.16}_{-0.14}$  \\
$\log_{10}\, r$        & $> -11.2 $ &$ -19.2\pm 3.5$ &  $ -7.34^{+0.62}_{-0.85}$& $> -3.36 $  \\
$ \log_{10}\, \left(T_{re} / \text{GeV}\right)$        & $ < 4.84$ & $ < -1.18$&  $- $& $ < -3.06$  \\
$ w_{re}$        & $ 0.36^{+0.15}_{-0.28} $ &$0 $ & $1/3$ & $ 1$  \\
$ x_{re}$        & $ -8.6^{+7.3}_{-3.3} $ &$-23.41^{+0.39}_{-1.1} $ & $ -7.34^{+0.62}_{-0.85}$ & $ 0.59\pm 0.72$  \\
\hline
\hline

\end{tabular}
}

 \caption{The mean $\pm 1 \sigma$ constraints on the primordial and reheating parameters inferred from the NANOGrav 15-yr data for different reheating models.  }
 \label{tab:CBSHD}
\end{table}

\begin{figure}[!ht]
    \centering
    \includegraphics[scale=0.7]{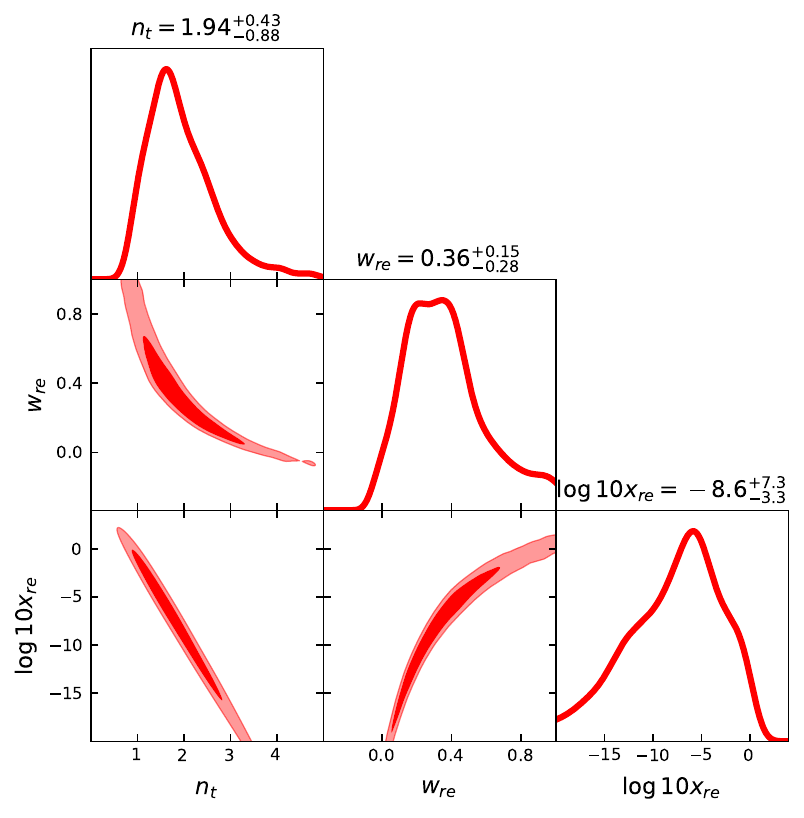}
    \caption{Posterior distributions of primordial and reheating parameters ($n_t, w_{re}$ and $x_{re}$) for arbitrary reheating eos.}
    \label{fig: reheatwre}
\end{figure}
  \begin{figure}[!ht]
    \centering
    \includegraphics[scale=0.7]{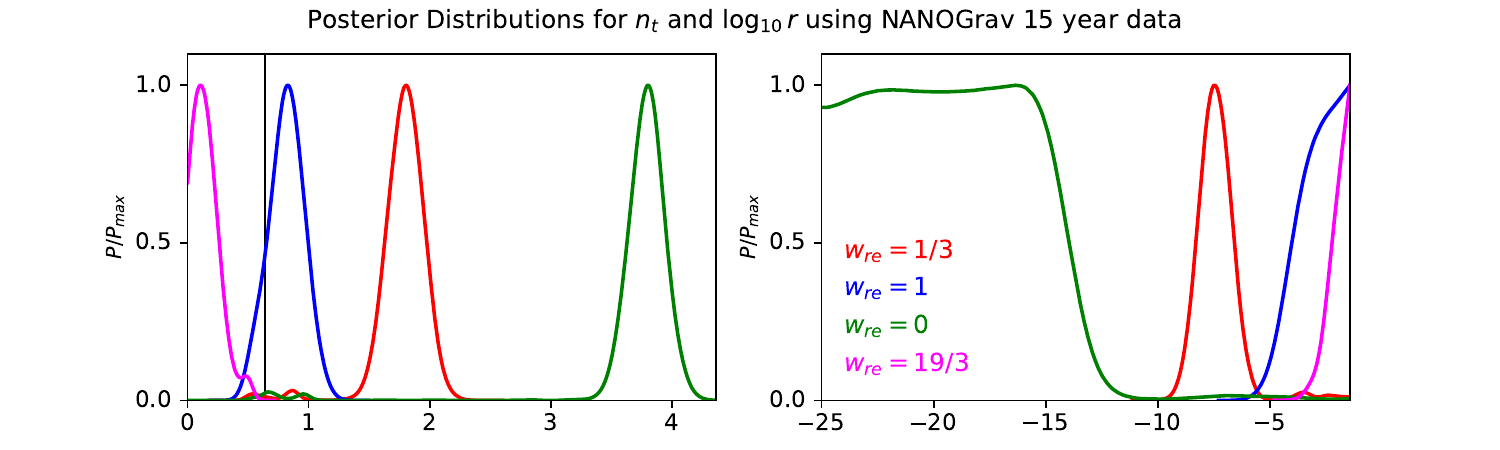}
    \caption{The 1-dimensional normalized posterior distributions for $n_t$ (left) and $\log_{10} r$ (right) for different models of reheating. The black solid line corresponds to the spectral index of SMBHBs ($n_t = 7 /11$).}
    \label{fig: reheatwr}
\end{figure}
Fig. (\ref{fig: reheatwre}) is the triangular plot showing the posterior distributions for $n_t$, $w_{re}$ and $x_{re}$ with $68 \%$ and $95\%$ Confidence Level (CL).  We find $n_t = 1.94^{+0.43}_{-0.86} $ and $w_{re} = 0.36^{+0.15}_{-0.28}$, a lower bound on $r$, and a maximal reheating temperature of $T_{re}\leq 10^5 \, GeV$ at 68\% CL, i.e. NANOGrav data prefers instantaneous reheating and a low reheating temperature. The inferred value of $n_t$ is linked to the best-fit power spectral density (PSD) spectral index $\gamma \approx 3.2$. The expected value of $\gamma = 13/3$ from the merging supermassive black-holes binaries (SMBHBs) corresponds to $n_T \approx 7 / 11$ which lies well outside the 95 $\%$ CL. Figure (\ref{fig: reheatwre}) illustrates the fact that increasing $w_{re}$ results in a smaller inferred value of $n_t$. %Furthermore, we find bounds on scalar to tensor ratio $r$ and reheating temperature $T_{re}$. We find negative correlations of $n_t$ with $w_{re}$ and $x_{re}$. 

Next, we consider the different values of $w_{re} = 0, 1/3 $ and $1$. The 68 $\%$ CL constraints for the above-mentioned models are tabulated in Table \ref{tab:CBSHD}. First, let us consider the canonical reheating scenario \cite{Dai:2014jja} which is defined by setting $w = 0$. In this scenario, we find $n_t = 3.72 ^{+0.23}_{-0.071}$ and $\log_{10} r =  -19.2 \pm 3.5$ at 1 $\sigma$ credible intervals. Furthermore, the upper bound on reheating temperature comes out to be $\log_{10}\, \left(T_{re} / \text{GeV}\right) < -1.18 $. Moving to the instantaneous reheating case, i.e., $w_{re} = 1/3$, we reproduce the results presented in \cite{Vagnozzi:2023lwo}. For the sake of completeness, we report that  $n_t = 1.77^{+0.18}_{-0.12}$ and $\log_{10} r = -7.34^{+0.62}_{-0.85}$. Finally, we discuss the stiff reheating scenario ($w_{re} = 1$). We report that the NANOGrav data points towards the lower tensor spectral index $n_t = 0.82 ^{+0.16}_{-0.14}$ and lower bound on tensor to scalar ratio $ \log_{10} r > -3.36$. The analysis of the latest 15-year NANOGrav data suggests that there is a strong correlation between the reheating eos $w_{re}$ and inferred $n_t$. As a consistency check, we performed the MCMC analysis with fixed $w_{re} = 19/3$, which corresponds to $n_t=0$ from \eqref{eq: GWInfl},\eqref{eq:gammaPTA} \footnote{ One can derive this specific equation of state using the eq. (\ref{eq:gammaPTA}) while assuming the $\gamma = 3.2$. However during the analysis, we do not fix the value of $\gamma = 3.2$ in our analysis to infer $n_t$ and $r$ } and as expected we find the resulting value of $n_t = 0.165^{+0.045}_{-0.15} $ and lower bound on $r>0.007$ which is much closer to the upper bound given from the joint analysis of Planck, WMAP, BICEP2, and BICEP3 \cite{BICEP:2021xfz}, $r<0.036$. We show the 1-dimensional normalized posterior distributions for $n_t$ and $\log_{10} r$ for different reheating models in Figure (\ref{fig: reheatwr}). The solid black line in Figure (\ref{fig: reheatwr}) corresponds to the $\gamma$ from SMBHBs. We have checked the robustness of our analysis for wide choices of priors of model parameters.

The 15-year NANOGrav PTA signal predicts the strong constraints on tensor spectral index $n_t$ and reheating eos $w_{re}$. Such a prediction of the blue spectrum and instantaneous reheating eos set strong implications for the predictions for $\Delta N_{eff}$ and violation of upper limits on the SGWB amplitudes on interferometer scales for the primordial parameters inferred from the PTA signal. To estimate the $\Delta N_{eff}$ contributions from PTA signal, we use the integral mentioned in eq. (\ref{eq:Neff}) at the BBN scales. This contribution depends on the choice of lower and upper limits appearing in eq. (\ref{eq:Neff}). At BBN scales, lower limit $f_{min}$ corresponds to the size of the comoving horizon at the time BBN which can be safely assumed to be $10^{-10}$Hz \cite{Smith:2006nka,Cabass:2015jwe}. The choice of upper limit $f_{max}$ is quite debatable, as it depends on the reheating temperature $T_{re}$. In this work, we find the $f_{max}$ such that $\Delta N_{eff}$ bounds from BBN are not violated from the inflationary and reheating parameters inferred from the NANOGrav data. We find the $f_{max}$ corresponding to $\Delta N_{eff} = 0.4$ for different reheating models discussed previously. For the arbitrary $w_{re}$, using the best-fit values of model parameters, we find that the allowed value is $f_{max} = 5 \times 10^{-6}$. Performing a similar exercise to other reheating models results in the allowed value of $f_{max} $ spanning from [$1.4 \times 10^{-7}, 1.5 \times 10^{-5}$]. Finally, if we assume instantaneous reheating and the same tilt, the most likely value of $r\sim 10^{-7}$ and a constant $n_t$ throughout the $62$ e-folds of (i.e. $f_{max}=10^{10}$ Hz), then $n_t<0.3$ in discord with the NANOGrav signal. We defer a more rigorous joint analysis to future work. 

A much-anticipated point related to the blue spectrum described by a power law is its extrapolation at least to LIGO scales. Such an extrapolation would violate the upper limits from the LIGO i.e. $\Omega_{GW} \leq 1.7 \times 10^{-8}$ at $f \sim 25$Hz \cite{KAGRA:2021kbb}. We find that the incorporation of the reheating phase (with constant eos) also runs into the same problems. However, in order to avoid such a problem, there are several models in the literature that gives a break in the power-law tensor power spectrum \cite{Allen:1996sw,Wang:2023ost,Benetti:2021uea,HosseiniMansoori:2023mqh,Cheung:2023ihl,Choudhury:2023kam}. In the section that follows we discuss a few ways to reconcile the GW interferometer limits (Absence of GW detection by LIGO) with our results from the analysis of NANOGrav data.
\section{NANOGrav in accordance with LIGO}
\label{sec:nli}
The analysis performed in section (\ref{sec: analysis}) will violate the bounds on GW spectrum from the LIGO \cite{KAGRA:2021kbb} and $\Delta N_{eff}$ if the primordial gravitational wave spectrum in eq. (\ref{eq: GWInfl}) is extrapolated to LIGO scales. We discuss two different methods to avoid this problem in the subsections below. Note that these methods are purely phenomenological and do not require a break in the spectrum, which should be based on serious considerations.

\subsection{Mixing of Astrophysical and Primordial Gravitational Wave Backgrounds} So far we have studied the observed SGWB from NANOGrav to be purely primordial. It could very well be that there exist several gravitational wave backgrounds (GWB) instead of only a primordial one. An obvious way to address the bounds from LIGO and other interferometer experiments is to envisage the existence of additional astrophysical background on top of the primordial GWB. It is a challenging task to write the gravitational wave energy spectrum associated with astrophysical processes as it depends upon several complex physical phenomena and lies beyond the scope of this paper. To account for the astrophysical GWB, we assume that the present-day astrophysical gravitational waves energy spectrum has the same frequency dependence as primordial GWB, for simplicity. Then the astrophysical gravitational wave energy spectrum (AGWES) today is described as 
\begin{eqnarray}
    \Omega_{\text{GW}}^{\text{Astro}} (f) &=& A^{\text{Astro}} \, \left(\frac{f}{f_{*}}\right)^{n_t + 2 \alpha}\,. \label{eq: Astro}
\end{eqnarray}
Here $A^{\text{Astro}}$ is the amplitude of AGWES. We remind the reader that such a choice of the spectrum is purely phenomenological and relevant only near the scales of PTA and becomes negligible at the interferometer scales as these astrophysical processes such as incoherent mergers are on specific scales, unlike the primordial spectrum. After taking into account the contribution from eq. (\ref{eq: Astro}) the total energy spectrum is expressed as 
\begin{eqnarray}
    \Omega_{\text{GW}}^{\text{PTA}} &=& \Omega_{\text{GW}}^{\text{Astro}} + \Omega_{\text{GW}}^{\text{prim}} \label{eq:astrotot}
\end{eqnarray}
It is straightforward to derive the functional form of amplitude and spectral index of PTA signal in terms of $A_{astro}$ and early universe parameters as done in section \ref{sec: pulsar}. We perform the MCMC analysis with additional parameter $A_{astro}$ along with the early universe parameters. We work with a fixed reheating temperature and as it has been shown in section \ref{sec: analysis} that there exists a degeneracy between $T_{re}$ and $r$ so fixing $T_{re}$ helps us to bound $r$ appropriately\footnote{Note that fixing the value reheating temperature $T_{re}$ does not affect the conclusions made in the analysis.}.

    \begin{table}[H]
    \centering
68 $\%$  Constraints from NANOGrav 15-year data considering Astrophysical and Primordial GWBs
   \vspace{1 em}  \\
{
\begin{tabular}{|c|c|c|c|c|}

\hline
\hline

    Parameter         & $w_{re}$ free &  $w_{re} = 0$ & $w_{re} = 1/3$ &  $ w_{re} = 1$ \\
\hline
\hline
$ n_t $        & $ 1.62^{+0.27}_{-0.91} $ & $3.76^{+0.19}_{-0.11} $& $ 1.80 \pm 0.15$ & $ 0.81\pm 0.14$  \\
$\log_{10}\, r$        & $< -12.0 $ &$ -29^{+11}_{-19}$ &  $ < -10.7$& $ \text{unconstrained}$  \\
$ \log_{10}\, \left(A^{\text{Astro}}_{R}\right)$        & $ 4.19^{+4.6}_{-0.63}$ & $ < 6.09$&  $4.15^{+4.6}_{-0.63} $& $ 4.43^{+4.4}_{-0.64}$  \\
$ w_{re}$        & $ \text{unconstrained} $ &$0 $ & $1/3$ & $ 1$  \\
\hline
\hline

\end{tabular}
}
 \caption{The mean $\pm 1 \sigma$ constraints on the model parameters inferred from the NANOGrav 15-yr data for different reheating models. Note that $A_{R}^{\text{Astro}}\equiv \frac{A^{\text{Astro}}}{\left(f_{*}\right)^{n_t + 2 \alpha}}$ is a rescaled amplitude of Astrophysical GW energy spectrum. }
 \label{tab:CBSHD1}
\end{table}
We present the 68 \% CL constraints considering the mix of Astrophysical and Primordial GWBs using the NANOGrav 15-year data for different reheating models in Table (\ref{tab:CBSHD1}). It is worth mentioning that inferred value of the tensor spectral index in this scenario remains pretty much unchanged. This is expected because the relation of $\gamma$ with $n_t$ and $w_{re}$ is the same as the purely primordial case. The re-scaled amplitude of AGWES, defined as $A_{R}^{\text{Astro}}\equiv A^{\text{Astro}} / \left(f_{*}\right)^{n_t + 2 \alpha} $, is constrained at 68 \% CL \footnote{We have used $f_*=f_{yr}$ in the analysis.}. We report $\log_{10} \, \left(A_{R}^{\text{Astro}}\right) = 4.19^{+4.8}_{-0.3}$ and $\log_{10} r < -12.0$ setting $w_{re}$ to be free. In this scenario $w_{re} $ is unconstrained. The results of different reheating eos are shown in Table (\ref{tab:CBSHD1}).

\begin{figure}[!ht]
    \centering
    \includegraphics[scale=0.7]{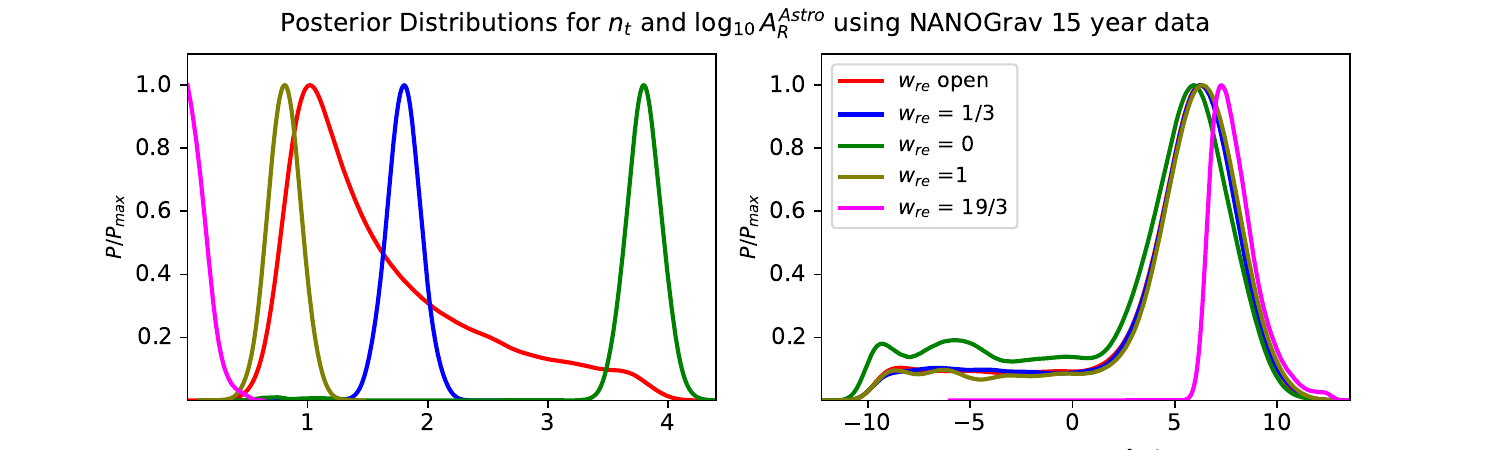}
    \caption{The 1-dimensional normalized posterior distributions for $n_t$ and $\log_{10} A^{Astro}_{R}$ for different models of reheating}
    \label{fig: mixgw}
\end{figure}

It is important to note that while considering  the mix of astrophysical and primordial GWBs, we find an upper bound on $\log_{10} r$ to be $ < -12.0$ and $< -10.7$ for free $w_{re}$ and instant reheating case respectively. For the canonical reheating scenario ($w_{re} = 0$) we find $\log_{10} r= -29^{+11}_{-19}$ at 1 $\sigma$ limits. These inferred values of tensor to scalar ratio play a significant role to satisfy the LIGO bound. One can do an independent calculation without accounting for the NANOGrav data to find the limiting value of $r$ for given $n_t$. For the blue spectrum as in our case, this limiting value turns out to be $\log_{10} r\approx -26$\footnote{This limit on $r$ is derived considering the results of \cite{KAGRA:2021kbb}}. Hence, for the extrapolated spectrum, the LIGO bound is still much stronger. %Taking this limit at its face value, we find that the constraints from our analysis are in accordance with the LIGO bounds. 

Using the inferred values of model parameters we can estimate the amount of primordial GWB with respect to the AGWES. We define the parameter 
 $\tilde \beta\equiv \Omega_{GW}^{prim}/\Omega_{GW}^{Astro}$ as the ratio of amplitudes of primordial GW and astrophysical GW spectrum. For the case of instant reheating case, using the parameter values at 68 \% from Table (\ref{tab:CBSHD1}), we find $\tilde \beta = 0.164$, meaning that around 14.09 \% of total PTA signal could be primordial. However, imposing the LIGO constraint i.e. $r \sim 10^{-26}$, we find that almost all PTA signal belongs to astrophysical. Finally, we show the 1-dimensional plots for the re-scaled amplitude of AGWES for different reheating eos in Fig. (\ref{fig: mixgw})

\begin{figure}[!ht]
    \centering
    \includegraphics[scale=0.8]{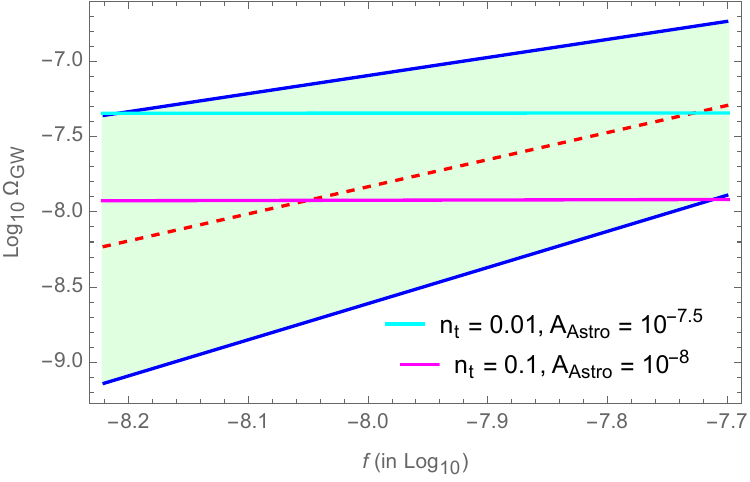}
    \caption{Evolution of GW energy spectrum for mean $ \pm 2\sigma$ values inferred from NANOGrav 15 year results. The two blue and one red-dashed lines show the gravitational waves energy spectrum for mean $ + 2\sigma$ , mean $ - 2\sigma$  and mean values respectively. The shaded green region is the allowed parameter space in accord with NANOGrav results. The cyan and magenta lines are the GW energy spectrum for parameter values respecting both NANOGrav and LIGO constraints. }
    \label{fig: ligo-nano}
\end{figure}

Next, we consider the mixing of constant AGWES present along with the primordial GWB. That is, we consider \eqref{eq:astrotot} but with $\Omega_{\text{GW}}^{\text{Astro}} =$ constant. We did not perform a Bayesian analysis as it lies beyond the scope of this article. To get a glimpse of allowed values of model parameters, we plot the \eqref{eq: GWPTA} with 2-$\sigma$ credible intervals as presented in Fig. (\ref{fig: ligo-nano}). We get that $n_t \in [0,0.1]$ and $\log_{10} A_{\text{Astro}} \in [-8,-7.5]$ lies within the NANOGrav band as well as respect the LIGO constraint. %We can further find the amount of primordial signal present with respect to the total PTA signal.
For the values illustrated in Fig. (\ref{fig: ligo-nano}), we find that the primordial signal is between 34 \% to 55 \% of the PTA SGWB signal for $n_t = 0 , \log_{10} A_{\text{Astro}} = -7.5$ and $n_t = 0.1 , \log_{10} A_{\text{Astro}} = -8.0$.
  \subsection{Running of the Spectrum}
\begin{figure}[!ht]
    \centering
    \includegraphics[scale=0.8]{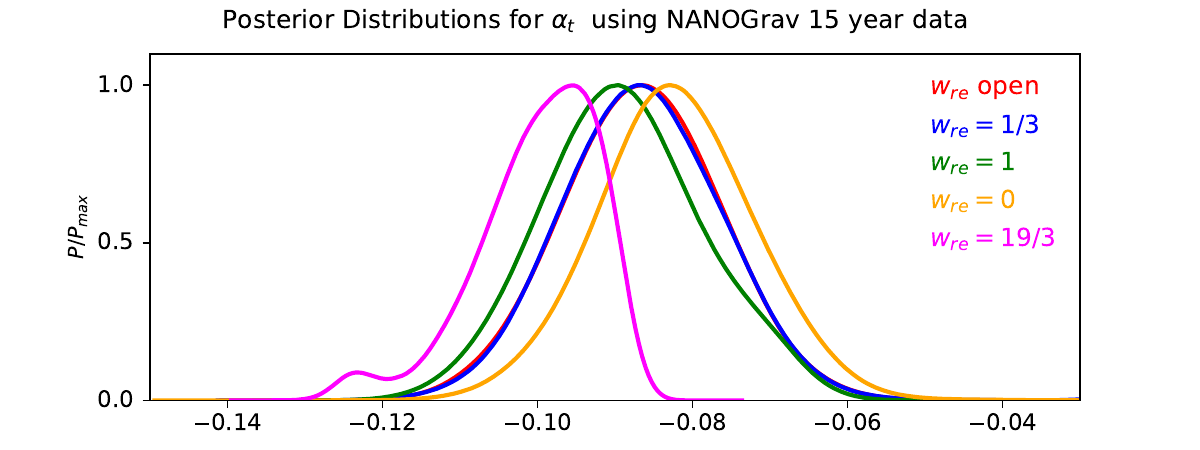}
    \caption{The 1-dimensional normalized posterior distributions for $\alpha_t$ for different models of reheating.}
    \label{fig: alpha}
\end{figure}

So far in the analysis, the spectrum of gravitational waves is assumed to have a constant spectral index, i.e. the spectral index is independent of the scale. The tensor spectrum of inflation typically has a running of the spectral index, albeit extremely small. The running of the spectrum for tensor perturbations $\alpha_t$ is defined as 
\begin{equation}
    \alpha_t = \dfrac{d^2\log \mathcal{P}^{\text{prim}}_T}{d\log f^2}|_{f=f_*}. 
    \label{eq:running}
\end{equation}
Typically, this factor is smaller than the spectral index by a factor approximately given by the number of e-folds. Given that the number of e-folds is sufficiently large, running is relatively insignificant. 
However, for such large $n_t$ (that we have obtained from our analysis of NANOGrav) such as in the case of instant reheating, the running is perhaps significant enough to drive the GW amplitude small enough to a point where it is unobserved by the interferometer experiments. In order to test this hypothesis we extrapolate the result in equation \eqref{eq: GWInfl} to LIGO scales ($f\sim 25$ Hz ). %, and require that the spectrum we obtain is lower than the current bound % expected minimum amplitude of the GW spectrum
%at these scales, $\Omega_{\text{GW}} \leq 1.7 \times 10^{-8}$ . 
The amplitude of the spectrum we obtain by extrapolating equation \eqref{eq: GWInfl} should satisfy the LIGO bound \cite{KAGRA:2021kbb},  expressed as
\begin{equation}
    \Omega_{\text{GW}}(f_{\text{\text{LIGO}}}) \leq 1.7 \times 10^{-8},
    \label{eq:gwligo}
\end{equation}
if we were to assume NANOGrav data is primordial GW background.
Assuming that running of the spectrum is significant (which is a reasonable assumption to make for large $n_t$), $n_t$ in the equation \eqref{eq: GWInfl} is modified as
\begin{equation}
    n_t(f) = n_t(f_{yr}) + \frac{\alpha_t}{2} \log\left( \frac{f}{f_{yr}}\right)
\end{equation}
where $f_{\text{yr}}$ is the frequency mode corresponding to the NANOGrav data ($f_{yr} = 1 yr^{-1} \sim 3.17 \times 10^{-8}$Hz). We solve equation \eqref{eq:gwligo} with parameters obtained from the likelihood analysis performed in section \ref{sec:Discussion}. The results of our analysis are demonstrated in the table (\ref{tab:alpha}), and the posterior distribution for the running parameter $\alpha_t$ in light of various reheating scenarios is presented in figure \ref{fig: alpha}.
    \begin{table}[H]
    \centering
68 $\%$  Constraints from NANOGrav 15-year data with constraints on running from LIGO bounds
   \vspace{1 em}  \\
{
\begin{tabular}{|c|c|c|c|c|}

\hline
\hline

    Parameter         & $w_{re}$ free &  $w_{re} = 0$ & $w_{re} = 1/3$ &  $ w_{re} = 1$ \\
\hline
\hline
$ n_t $        & $ 1.94^{+0.43}_{-0.88} $ & $3.72^{+0.23}_{-0.071} $& $ 1.77^{+0.18}_{-0.12}$ & $ 0.82^{+0.16}_{-0.14}$  \\
$\log_{10}\, r$        & $> -11.2 $ &$ -19.2\pm 3.5$ &  $ -7.34^{+0.62}_{-0.85}$& $> -3.36 $  \\
$ x_{re}$        & $ -8.6^{+7.3}_{-3.3} $ &$-23.41^{+0.39}_{-1.1} $ & $ -7.34^{+0.62}_{-0.85}$ & $ 0.59\pm 0.72$   \\
$ \log_{10}\, \left(T_{re} / \text{GeV}\right)$        & $ < 4.78$ & $ < -1.16$&  $- $& $ < -3.09$  \\
$ w_{re}$        & $ 0.36^{+0.15}_{-0.28} $ &$0 $ & $1/3$ & $ 1$  \\
$\alpha_t$ & $ -0.087\pm 0.010 $ &$ -0.0769^{+0.0051}_{-0.016} $ & $ -0.0847^{+0.0087}_{-0.013}$ & $-0.0890^{+0.0099}_{-0.011}$  \\
\hline
\hline

\end{tabular}
}
 
 \caption{The mean $\pm 1 \sigma$ constraints on the primordial and reheating parameters with constraints on running of the spectrum obtained from LIGO bounds (last row of the table).}
 \label{tab:alpha}
\end{table}
For instantaneous reheating, if the running of the spectrum has a value $\alpha_t=-0.0847^{+0.0087}_{-0.013}$ we do not violate the bounds set by LIGO. Similarly, for our best-fit parameters, this value of running is $\alpha_t= -0.087\pm 0.010 $ for $w_{re}$ free, and $ -0.0769^{+0.0051}_{-0.016}$ for canonical reheating with $w_{re}=0$. Interestingly we find that irrespective of the reheating model $\alpha_t\sim -0.08$ can fit the NANOGrav data without violating the LIGO bounds. 
\section{Conclusions and Future directions}
\label{sec: conclusions}
We have analyzed the NANOGrav 15-year dataset to constrain the early universe parameters in the presence of a nontrivial reheating phase. We parameterized the power spectrum of GW as a power law. Reheating phase is characterized by an equation of state $w_{re}$ and the duration of the reheating phase is determined by $T_{re}$, i.e. temperature at the end of the reheating phase. In our analysis we find a mean value of $w_{re} = 0.36\pm 0.2$ which is very close to the equation of state of the radiation domination era, $w_{rad} = 1/3$, suggesting that at the end of inflation, the universe transitions immediately to a radiation-dominated era. While our analysis is suggestive of instantaneous reheating methods over slower reheating processes, slow reheating processes are fine as well. The sizeable error bars in $w_{re}$ can accommodate alternative reheating mechanisms characterized by both stiffer ($w_{re}>1/3$) and milder ($w_{re}<1/3$) equations of state.
The corresponding value of $n_t$ is $1.94^{+0.43}_{-0.86}$ at 68 \%  CL, and the reheating temperature is highly limited $T_{re}\leq 10^5 \, GeV$. 

Our analysis suggests a strong correlation between the reheating eos and the inferred spectral index from the NANOGrav data. First, we have confirmed the results of previous studies for instant reheating. Second, we have shown that the reheating phase with stiff matter can drive down the value of the tensor spectral index of primordial gravitational waves inferred from the NANOGrav data to near-scale invariance.  Finally, we infer that in general, NANOGrav dataset tends to produce a smaller blue tilt for larger values of the reheating parameter $w_{re}$. Kinetic reheating for example points toward a tensor spectral index of $n_t = 0.82 ^{+0.16}_{-0.14}$ and lower bound on tensor to scalar ratio $ \log_{10} r > -3.36$, while scale invariance $n_t$ requires a larger value of $w_{re} = 19/3$. Our likelihood analysis confirmed this result for this value of $w_{re}$ with $n_t = 0.165^{+0.045}_{-0.15} $, i.e. scale invariance is 1$\sigma$ away. The best-fit result for our data is highly blue-tilted spectrum with $n_t \sim 2$ similar to bounce models. Single-field bounce scenarios are usually ruled out due to the blue-tilted scalar spectrum. However, in a series of articles pertaining to a paradigm of a sourced bounce \cite{Ben-Dayan:2016iks,Ben-Dayan:2018ksd,Artymowski:2020pci,Ben-Dayan:2023rlj} the authors have pointed out this is circumvented by introducing a gauge field that sources the perturbations. In the scenario of sourced bounce the vacuum spectrum will remain blue-tilted. It will be interesting to examine the possibility of the NANOGrav signal being related to the vacuum spectrum of sourced bounce.

As we pointed out earlier the spectrum from NANOGrav is inconsistent with the upper limits from LIGO. We hypothesize that this issue can be solved in several ways. We have primarily analyzed two methods that help us resolve the inconsistency of signal with LIGO bounds. In the first case, we propose that the NANOGrav signal could be a mixture of several stochastic GW backgrounds including the primordial GW spectrum. For the sake of simplicity, we have assumed the same scale dependence for GWs from astrophysical phenomena as that of the primordial spectrum or that the astrophysical background is a simple constant background. In the first method, the simple analysis tells us that for the LIGO bound to be respected, the percentage of the GW spectrum that is primordial cannot be larger than 16 percent. We propose that a detailed study should be undertaken in order to distinguish astrophysical sources from primordial GWs. The constant amplitude method allowed the primordial signal to be more than $34\%$ of the total signal, and in accord with the LIGO bound, but in turn allowed only a small blue tilt at the level of $n_t<0.1$. Another proposal independent of resolving the spectrum into components has to do with a significant running of the spectrum. Our calculations show that a running of $\alpha_t\sim -0.08$ makes sure that the extrapolation of the primordial signal (with parameters obtained from NANOGrav) will respect the LIGO bounds.

We have clearly shown that the LIGO bounds can be reconciled with NANOGrav signal being primordial. However, a significant amount of work is required in this direction. We hope that this analysis has convinced the reader that it is too early to rule out the primordial origins of this signal. During our analysis, we assumed that the reheating phase has a constant equation of state. This need not be true. Time and frequency dependence of the reheating parameters could play a role in altering the spectral index of GWs. Further analysis is required to fully understand the effect of reheating phase on GW spectrum at NANOGrav scales. Preferably such analysis should combine the different datasets of 
PTA, CMB and LIGO all together and including $N_{eff}$ constraints. Assuming the primordial origins of the signal, we suggest a reheating mechanism that involves stiff matter as a way to reconcile it with the scale invariant GW spectrum of standard canonical single field slow-roll inflation. We save interesting questions such as resolving the spectrum and analyzing the details of the reheating phase for future works. 

\section*{Acknowledgements}
We acknowledge the Ariel HPC Center at Ariel University for providing computing resources that have contributed to the research results reports reported within this paper.
\bibliography{reference}

\begin{thebibliography}{104}%
\makeatletter
\providecommand \@ifxundefined [1]{%
 \@ifx{#1\undefined}
}%
\providecommand \@ifnum [1]{%
 \ifnum #1\expandafter \@firstoftwo
 \else \expandafter \@secondoftwo
 \fi
}%
\providecommand \@ifx [1]{%
 \ifx #1\expandafter \@firstoftwo
 \else \expandafter \@secondoftwo
 \fi
}%
\providecommand \natexlab [1]{#1}%
\providecommand \enquote  [1]{``#1''}%
\providecommand \bibnamefont  [1]{#1}%
\providecommand \bibfnamefont [1]{#1}%
\providecommand \citenamefont [1]{#1}%
\providecommand \href@noop [0]{\@secondoftwo}%
\providecommand \href [0]{\begingroup \@sanitize@url \@href}%
\providecommand \@href[1]{\@@startlink{#1}\@@href}%
\providecommand \@@href[1]{\endgroup#1\@@endlink}%
\providecommand \@sanitize@url [0]{\catcode `\\12\catcode `\$12\catcode
  `\&12\catcode `\#12\catcode `\^12\catcode `\_12\catcode `\%12\relax}%
\providecommand \@@startlink[1]{}%
\providecommand \@@endlink[0]{}%
\providecommand \url  [0]{\begingroup\@sanitize@url \@url }%
\providecommand \@url [1]{\endgroup\@href {#1}{\urlprefix }}%
\providecommand \urlprefix  [0]{URL }%
\providecommand \Eprint [0]{\href }%
\providecommand \doibase [0]{http://dx.doi.org/}%
\providecommand \selectlanguage [0]{\@gobble}%
\providecommand \bibinfo  [0]{\@secondoftwo}%
\providecommand \bibfield  [0]{\@secondoftwo}%
\providecommand \translation [1]{[#1]}%
\providecommand \BibitemOpen [0]{}%
\providecommand \bibitemStop [0]{}%
\providecommand \bibitemNoStop [0]{.\EOS\space}%
\providecommand \EOS [0]{\spacefactor3000\relax}%
\providecommand \BibitemShut  [1]{\csname bibitem#1\endcsname}%
\let\auto@bib@innerbib\@empty
%</preamble>
\bibitem [{\citenamefont {Caprini}\ and\ \citenamefont
  {Figueroa}(2018)}]{Caprini:2018mtu}%
  \BibitemOpen
  \bibfield  {author} {\bibinfo {author} {\bibfnamefont {Chiara}\ \bibnamefont
  {Caprini}}\ and\ \bibinfo {author} {\bibfnamefont {Daniel~G.}\ \bibnamefont
  {Figueroa}},\ }\bibfield  {title} {\enquote {\bibinfo {title} {{Cosmological
  Backgrounds of Gravitational Waves}},}\ }\href {\doibase
  10.1088/1361-6382/aac608} {\bibfield  {journal} {\bibinfo  {journal} {Class.
  Quant. Grav.}\ }\textbf {\bibinfo {volume} {35}},\ \bibinfo {pages} {163001}
  (\bibinfo {year} {2018})},\ \Eprint {http://arxiv.org/abs/1801.04268}
  {arXiv:1801.04268 [astro-ph.CO]} \BibitemShut {NoStop}%
\bibitem [{\citenamefont {Renzini}\ \emph {et~al.}(2022)\citenamefont
  {Renzini}, \citenamefont {Goncharov}, \citenamefont {Jenkins},\ and\
  \citenamefont {Meyers}}]{Renzini:2022alw}%
  \BibitemOpen
  \bibfield  {author} {\bibinfo {author} {\bibfnamefont {Arianna~I.}\
  \bibnamefont {Renzini}}, \bibinfo {author} {\bibfnamefont {Boris}\
  \bibnamefont {Goncharov}}, \bibinfo {author} {\bibfnamefont {Alexander~C.}\
  \bibnamefont {Jenkins}}, \ and\ \bibinfo {author} {\bibfnamefont {Pat~M.}\
  \bibnamefont {Meyers}},\ }\bibfield  {title} {\enquote {\bibinfo {title}
  {{Stochastic Gravitational-Wave Backgrounds: Current Detection Efforts and
  Future Prospects}},}\ }\href {\doibase 10.3390/galaxies10010034} {\bibfield
  {journal} {\bibinfo  {journal} {Galaxies}\ }\textbf {\bibinfo {volume}
  {10}},\ \bibinfo {pages} {34} (\bibinfo {year} {2022})},\ \Eprint
  {http://arxiv.org/abs/2202.00178} {arXiv:2202.00178 [gr-qc]} \BibitemShut
  {NoStop}%
\bibitem [{\citenamefont {Hobbs}\ and\ \citenamefont {Dai}(2017)}]{Hobbs_2017}%
  \BibitemOpen
  \bibfield  {author} {\bibinfo {author} {\bibfnamefont {George}\ \bibnamefont
  {Hobbs}}\ and\ \bibinfo {author} {\bibfnamefont {Shi}\ \bibnamefont {Dai}},\
  }\bibfield  {title} {\enquote {\bibinfo {title} {Gravitational wave research
  using pulsar timing arrays},}\ }\href {\doibase 10.1093/nsr/nwx126}
  {\bibfield  {journal} {\bibinfo  {journal} {National Science Review}\
  }\textbf {\bibinfo {volume} {4}},\ \bibinfo {pages} {707--717} (\bibinfo
  {year} {2017})}\BibitemShut {NoStop}%
\bibitem [{\citenamefont {Arzoumanian}\ \emph {et~al.}(2020)\citenamefont
  {Arzoumanian} \emph {et~al.}}]{NANOGrav:2020bcs}%
  \BibitemOpen
  \bibfield  {author} {\bibinfo {author} {\bibfnamefont {Zaven}\ \bibnamefont
  {Arzoumanian}} \emph {et~al.} (\bibinfo {collaboration} {NANOGrav}),\
  }\bibfield  {title} {\enquote {\bibinfo {title} {{The NANOGrav 12.5 yr Data
  Set: Search for an Isotropic Stochastic Gravitational-wave Background}},}\
  }\href {\doibase 10.3847/2041-8213/abd401} {\bibfield  {journal} {\bibinfo
  {journal} {Astrophys. J. Lett.}\ }\textbf {\bibinfo {volume} {905}},\
  \bibinfo {pages} {L34} (\bibinfo {year} {2020})},\ \Eprint
  {http://arxiv.org/abs/2009.04496} {arXiv:2009.04496 [astro-ph.HE]}
  \BibitemShut {NoStop}%
\bibitem [{\citenamefont {Goncharov}\ \emph {et~al.}(2021)\citenamefont
  {Goncharov} \emph {et~al.}}]{Goncharov:2021oub}%
  \BibitemOpen
  \bibfield  {author} {\bibinfo {author} {\bibfnamefont {Boris}\ \bibnamefont
  {Goncharov}} \emph {et~al.},\ }\bibfield  {title} {\enquote {\bibinfo {title}
  {{On the Evidence for a Common-spectrum Process in the Search for the
  Nanohertz Gravitational-wave Background with the Parkes Pulsar Timing
  Array}},}\ }\href {\doibase 10.3847/2041-8213/ac17f4} {\bibfield  {journal}
  {\bibinfo  {journal} {Astrophys. J. Lett.}\ }\textbf {\bibinfo {volume}
  {917}},\ \bibinfo {pages} {L19} (\bibinfo {year} {2021})},\ \Eprint
  {http://arxiv.org/abs/2107.12112} {arXiv:2107.12112 [astro-ph.HE]}
  \BibitemShut {NoStop}%
\bibitem [{\citenamefont {Chen}\ \emph {et~al.}(2021)\citenamefont {Chen} \emph
  {et~al.}}]{Chen:2021rqp}%
  \BibitemOpen
  \bibfield  {author} {\bibinfo {author} {\bibfnamefont {S.}~\bibnamefont
  {Chen}} \emph {et~al.},\ }\bibfield  {title} {\enquote {\bibinfo {title}
  {{Common-red-signal analysis with 24-yr high-precision timing of the European
  Pulsar Timing Array: inferences in the stochastic gravitational-wave
  background search}},}\ }\href {\doibase 10.1093/mnras/stab2833} {\bibfield
  {journal} {\bibinfo  {journal} {Mon. Not. Roy. Astron. Soc.}\ }\textbf
  {\bibinfo {volume} {508}},\ \bibinfo {pages} {4970--4993} (\bibinfo {year}
  {2021})},\ \Eprint {http://arxiv.org/abs/2110.13184} {arXiv:2110.13184
  [astro-ph.HE]} \BibitemShut {NoStop}%
\bibitem [{\citenamefont {Antoniadis}\ \emph {et~al.}(2022)\citenamefont
  {Antoniadis} \emph {et~al.}}]{Antoniadis:2022pcn}%
  \BibitemOpen
  \bibfield  {author} {\bibinfo {author} {\bibfnamefont {J.}~\bibnamefont
  {Antoniadis}} \emph {et~al.},\ }\bibfield  {title} {\enquote {\bibinfo
  {title} {{The International Pulsar Timing Array second data release: Search
  for an isotropic gravitational wave background}},}\ }\href {\doibase
  10.1093/mnras/stab3418} {\bibfield  {journal} {\bibinfo  {journal} {Mon. Not.
  Roy. Astron. Soc.}\ }\textbf {\bibinfo {volume} {510}},\ \bibinfo {pages}
  {4873--4887} (\bibinfo {year} {2022})},\ \Eprint
  {http://arxiv.org/abs/2201.03980} {arXiv:2201.03980 [astro-ph.HE]}
  \BibitemShut {NoStop}%
\bibitem [{\citenamefont {Agazie}\ \emph
  {et~al.}(2023{\natexlab{a}})\citenamefont {Agazie} \emph
  {et~al.}}]{NANOGrav:2023gor}%
  \BibitemOpen
  \bibfield  {author} {\bibinfo {author} {\bibfnamefont {Gabriella}\
  \bibnamefont {Agazie}} \emph {et~al.} (\bibinfo {collaboration} {NANOGrav}),\
  }\bibfield  {title} {\enquote {\bibinfo {title} {{The NANOGrav 15 yr Data
  Set: Evidence for a Gravitational-wave Background}},}\ }\href {\doibase
  10.3847/2041-8213/acdac6} {\bibfield  {journal} {\bibinfo  {journal}
  {Astrophys. J. Lett.}\ }\textbf {\bibinfo {volume} {951}},\ \bibinfo {pages}
  {L8} (\bibinfo {year} {2023}{\natexlab{a}})},\ \Eprint
  {http://arxiv.org/abs/2306.16213} {arXiv:2306.16213 [astro-ph.HE]}
  \BibitemShut {NoStop}%
\bibitem [{\citenamefont {Antoniadis}\ \emph
  {et~al.}(2023{\natexlab{a}})\citenamefont {Antoniadis} \emph
  {et~al.}}]{Antoniadis:2023rey}%
  \BibitemOpen
  \bibfield  {author} {\bibinfo {author} {\bibfnamefont {J.}~\bibnamefont
  {Antoniadis}} \emph {et~al.},\ }\bibfield  {title} {\enquote {\bibinfo
  {title} {{The second data release from the European Pulsar Timing Array III.
  Search for gravitational wave signals}},}\ }\href@noop {} {\bibfield
  {journal} {\bibinfo  {journal} {.}\ } (\bibinfo {year}
  {2023}{\natexlab{a}})},\ \Eprint {http://arxiv.org/abs/2306.16214}
  {arXiv:2306.16214 [astro-ph.HE]} \BibitemShut {NoStop}%
\bibitem [{\citenamefont {Reardon}\ \emph
  {et~al.}(2023{\natexlab{a}})\citenamefont {Reardon} \emph
  {et~al.}}]{Reardon:2023gzh}%
  \BibitemOpen
  \bibfield  {author} {\bibinfo {author} {\bibfnamefont {Daniel~J.}\
  \bibnamefont {Reardon}} \emph {et~al.},\ }\bibfield  {title} {\enquote
  {\bibinfo {title} {{Search for an Isotropic Gravitational-wave Background
  with the Parkes Pulsar Timing Array}},}\ }\href {\doibase
  10.3847/2041-8213/acdd02} {\bibfield  {journal} {\bibinfo  {journal}
  {Astrophys. J. Lett.}\ }\textbf {\bibinfo {volume} {951}},\ \bibinfo {pages}
  {L6} (\bibinfo {year} {2023}{\natexlab{a}})},\ \Eprint
  {http://arxiv.org/abs/2306.16215} {arXiv:2306.16215 [astro-ph.HE]}
  \BibitemShut {NoStop}%
\bibitem [{\citenamefont {Xu}\ \emph {et~al.}(2023)\citenamefont {Xu} \emph
  {et~al.}}]{Xu:2023wog}%
  \BibitemOpen
  \bibfield  {author} {\bibinfo {author} {\bibfnamefont {Heng}\ \bibnamefont
  {Xu}} \emph {et~al.},\ }\bibfield  {title} {\enquote {\bibinfo {title}
  {{Searching for the Nano-Hertz Stochastic Gravitational Wave Background with
  the Chinese Pulsar Timing Array Data Release I}},}\ }\href {\doibase
  10.1088/1674-4527/acdfa5} {\bibfield  {journal} {\bibinfo  {journal} {Res.
  Astron. Astrophys.}\ }\textbf {\bibinfo {volume} {23}},\ \bibinfo {pages}
  {075024} (\bibinfo {year} {2023})},\ \Eprint
  {http://arxiv.org/abs/2306.16216} {arXiv:2306.16216 [astro-ph.HE]}
  \BibitemShut {NoStop}%
\bibitem [{\citenamefont {Agazie}\ \emph
  {et~al.}(2023{\natexlab{b}})\citenamefont {Agazie} \emph
  {et~al.}}]{NANOGrav:2023hde}%
  \BibitemOpen
  \bibfield  {author} {\bibinfo {author} {\bibfnamefont {Gabriella}\
  \bibnamefont {Agazie}} \emph {et~al.} (\bibinfo {collaboration} {NANOGrav}),\
  }\bibfield  {title} {\enquote {\bibinfo {title} {{The NANOGrav 15 yr Data
  Set: Observations and Timing of 68 Millisecond Pulsars}},}\ }\href {\doibase
  10.3847/2041-8213/acda9a} {\bibfield  {journal} {\bibinfo  {journal}
  {Astrophys. J. Lett.}\ }\textbf {\bibinfo {volume} {951}},\ \bibinfo {pages}
  {L9} (\bibinfo {year} {2023}{\natexlab{b}})},\ \Eprint
  {http://arxiv.org/abs/2306.16217} {arXiv:2306.16217 [astro-ph.HE]}
  \BibitemShut {NoStop}%
\bibitem [{\citenamefont {Agazie}\ \emph
  {et~al.}(2023{\natexlab{c}})\citenamefont {Agazie}, \citenamefont
  {Anumarlapudi}, \citenamefont {Archibald}, \citenamefont {Baker},
  \citenamefont {B{\'e}csy}, \citenamefont {Blecha}, \citenamefont {Bonilla},
  \citenamefont {Brazier}, \citenamefont {Brook}, \citenamefont {Burke-Spolaor}
  \emph {et~al.}}]{agazie2023nanograv}%
  \BibitemOpen
  \bibfield  {author} {\bibinfo {author} {\bibfnamefont {Gabriella}\
  \bibnamefont {Agazie}}, \bibinfo {author} {\bibfnamefont {Akash}\
  \bibnamefont {Anumarlapudi}}, \bibinfo {author} {\bibfnamefont {Anne~M}\
  \bibnamefont {Archibald}}, \bibinfo {author} {\bibfnamefont {Paul~T}\
  \bibnamefont {Baker}}, \bibinfo {author} {\bibfnamefont {Bence}\ \bibnamefont
  {B{\'e}csy}}, \bibinfo {author} {\bibfnamefont {Laura}\ \bibnamefont
  {Blecha}}, \bibinfo {author} {\bibfnamefont {Alexander}\ \bibnamefont
  {Bonilla}}, \bibinfo {author} {\bibfnamefont {Adam}\ \bibnamefont {Brazier}},
  \bibinfo {author} {\bibfnamefont {Paul~R}\ \bibnamefont {Brook}}, \bibinfo
  {author} {\bibfnamefont {Sarah}\ \bibnamefont {Burke-Spolaor}},  \emph
  {et~al.},\ }\bibfield  {title} {\enquote {\bibinfo {title} {The nanograv
  15-year data set: Constraints on supermassive black hole binaries from the
  gravitational wave background},}\ }\href@noop {} {\bibfield  {journal}
  {\bibinfo  {journal} {arXiv preprint arXiv:2306.16220}\ } (\bibinfo {year}
  {2023}{\natexlab{c}})}\BibitemShut {NoStop}%
\bibitem [{\citenamefont {Agazie}\ \emph
  {et~al.}(2023{\natexlab{d}})\citenamefont {Agazie} \emph
  {et~al.}}]{NANOGrav:2023tcn}%
  \BibitemOpen
  \bibfield  {author} {\bibinfo {author} {\bibfnamefont {Gabriella}\
  \bibnamefont {Agazie}} \emph {et~al.} (\bibinfo {collaboration} {NANOGrav}),\
  }\bibfield  {title} {\enquote {\bibinfo {title} {{The NANOGrav 15-year Data
  Set: Search for Anisotropy in the Gravitational-Wave Background}},}\
  }\href@noop {} {\bibfield  {journal} {\bibinfo  {journal} {.}\ } (\bibinfo
  {year} {2023}{\natexlab{d}})},\ \Eprint {http://arxiv.org/abs/2306.16221}
  {arXiv:2306.16221 [astro-ph.HE]} \BibitemShut {NoStop}%
\bibitem [{\citenamefont {Afzal}\ \emph {et~al.}(2023)\citenamefont {Afzal}
  \emph {et~al.}}]{NANOGrav:2023hvm}%
  \BibitemOpen
  \bibfield  {author} {\bibinfo {author} {\bibfnamefont {Adeela}\ \bibnamefont
  {Afzal}} \emph {et~al.} (\bibinfo {collaboration} {NANOGrav}),\ }\bibfield
  {title} {\enquote {\bibinfo {title} {{The NANOGrav 15 yr Data Set: Search for
  Signals from New Physics}},}\ }\href {\doibase 10.3847/2041-8213/acdc91}
  {\bibfield  {journal} {\bibinfo  {journal} {Astrophys. J. Lett.}\ }\textbf
  {\bibinfo {volume} {951}},\ \bibinfo {pages} {L11} (\bibinfo {year}
  {2023})},\ \Eprint {http://arxiv.org/abs/2306.16219} {arXiv:2306.16219
  [astro-ph.HE]} \BibitemShut {NoStop}%
\bibitem [{\citenamefont {Agazie}\ \emph
  {et~al.}(2023{\natexlab{e}})\citenamefont {Agazie} \emph
  {et~al.}}]{NANOGrav:2023ctt}%
  \BibitemOpen
  \bibfield  {author} {\bibinfo {author} {\bibfnamefont {Gabriella}\
  \bibnamefont {Agazie}} \emph {et~al.} (\bibinfo {collaboration} {NANOGrav}),\
  }\bibfield  {title} {\enquote {\bibinfo {title} {{The NANOGrav 15 yr Data
  Set: Detector Characterization and Noise Budget}},}\ }\href {\doibase
  10.3847/2041-8213/acda88} {\bibfield  {journal} {\bibinfo  {journal}
  {Astrophys. J. Lett.}\ }\textbf {\bibinfo {volume} {951}},\ \bibinfo {pages}
  {L10} (\bibinfo {year} {2023}{\natexlab{e}})},\ \Eprint
  {http://arxiv.org/abs/2306.16218} {arXiv:2306.16218 [astro-ph.HE]}
  \BibitemShut {NoStop}%
\bibitem [{\citenamefont {Johnson}\ \emph {et~al.}(2023)\citenamefont
  {Johnson}, \citenamefont {Meyers}, \citenamefont {Baker}, \citenamefont
  {Cornish}, \citenamefont {Hazboun}, \citenamefont {Littenberg}, \citenamefont
  {Romano}, \citenamefont {Taylor}, \citenamefont {Vallisneri}, \citenamefont
  {Vigeland} \emph {et~al.}}]{johnson2023nanograv}%
  \BibitemOpen
  \bibfield  {author} {\bibinfo {author} {\bibfnamefont {Aaron~D}\ \bibnamefont
  {Johnson}}, \bibinfo {author} {\bibfnamefont {Patrick~M}\ \bibnamefont
  {Meyers}}, \bibinfo {author} {\bibfnamefont {Paul~T}\ \bibnamefont {Baker}},
  \bibinfo {author} {\bibfnamefont {Neil~J}\ \bibnamefont {Cornish}}, \bibinfo
  {author} {\bibfnamefont {Jeffrey~S}\ \bibnamefont {Hazboun}}, \bibinfo
  {author} {\bibfnamefont {Tyson~B}\ \bibnamefont {Littenberg}}, \bibinfo
  {author} {\bibfnamefont {Joseph~D}\ \bibnamefont {Romano}}, \bibinfo {author}
  {\bibfnamefont {Stephen~R}\ \bibnamefont {Taylor}}, \bibinfo {author}
  {\bibfnamefont {Michele}\ \bibnamefont {Vallisneri}}, \bibinfo {author}
  {\bibfnamefont {Sarah~J}\ \bibnamefont {Vigeland}},  \emph {et~al.},\
  }\bibfield  {title} {\enquote {\bibinfo {title} {The nanograv 15-year
  gravitational-wave background analysis pipeline},}\ }\href@noop {} {\bibfield
   {journal} {\bibinfo  {journal} {arXiv preprint arXiv:2306.16223}\ }
  (\bibinfo {year} {2023})}\BibitemShut {NoStop}%
\bibitem [{\citenamefont {Agazie}\ \emph
  {et~al.}(2023{\natexlab{f}})\citenamefont {Agazie} \emph
  {et~al.}}]{NANOGrav:2023pdq}%
  \BibitemOpen
  \bibfield  {author} {\bibinfo {author} {\bibfnamefont {Gabriella}\
  \bibnamefont {Agazie}} \emph {et~al.} (\bibinfo {collaboration} {NANOGrav}),\
  }\bibfield  {title} {\enquote {\bibinfo {title} {{The NANOGrav 15 yr Data
  Set: Bayesian Limits on Gravitational Waves from Individual Supermassive
  Black Hole Binaries}},}\ }\href {\doibase 10.3847/2041-8213/ace18a}
  {\bibfield  {journal} {\bibinfo  {journal} {Astrophys. J. Lett.}\ }\textbf
  {\bibinfo {volume} {951}},\ \bibinfo {pages} {L50} (\bibinfo {year}
  {2023}{\natexlab{f}})},\ \Eprint {http://arxiv.org/abs/2306.16222}
  {arXiv:2306.16222 [astro-ph.HE]} \BibitemShut {NoStop}%
\bibitem [{\citenamefont {Antoniadis}\ \emph
  {et~al.}(2023{\natexlab{b}})\citenamefont {Antoniadis} \emph
  {et~al.}}]{Antoniadis:2023puu}%
  \BibitemOpen
  \bibfield  {author} {\bibinfo {author} {\bibfnamefont {J.}~\bibnamefont
  {Antoniadis}} \emph {et~al.},\ }\bibfield  {title} {\enquote {\bibinfo
  {title} {{The second data release from the European Pulsar Timing Array II.
  Customised pulsar noise models for spatially correlated gravitational
  waves}},}\ }\href@noop {} {\bibfield  {journal} {\bibinfo  {journal} {.}\ }
  (\bibinfo {year} {2023}{\natexlab{b}})},\ \Eprint
  {http://arxiv.org/abs/2306.16225} {arXiv:2306.16225 [astro-ph.HE]}
  \BibitemShut {NoStop}%
\bibitem [{\citenamefont {Antoniadis}\ \emph
  {et~al.}(2023{\natexlab{c}})\citenamefont {Antoniadis} \emph
  {et~al.}}]{Antoniadis:2023aac}%
  \BibitemOpen
  \bibfield  {author} {\bibinfo {author} {\bibfnamefont {J.}~\bibnamefont
  {Antoniadis}} \emph {et~al.},\ }\bibfield  {title} {\enquote {\bibinfo
  {title} {{The second data release from the European Pulsar Timing Array IV.
  Search for continuous gravitational wave signals}},}\ }\href@noop {}
  {\bibfield  {journal} {\bibinfo  {journal} {.}\ } (\bibinfo {year}
  {2023}{\natexlab{c}})},\ \Eprint {http://arxiv.org/abs/2306.16226}
  {arXiv:2306.16226 [astro-ph.HE]} \BibitemShut {NoStop}%
\bibitem [{\citenamefont {Antoniadis}\ \emph
  {et~al.}(2023{\natexlab{d}})\citenamefont {Antoniadis} \emph
  {et~al.}}]{Antoniadis:2023xlr}%
  \BibitemOpen
  \bibfield  {author} {\bibinfo {author} {\bibfnamefont {J.}~\bibnamefont
  {Antoniadis}} \emph {et~al.},\ }\bibfield  {title} {\enquote {\bibinfo
  {title} {{The second data release from the European Pulsar Timing Array: V.
  Implications for massive black holes, dark matter and the early Universe}},}\
  }\href@noop {} {\bibfield  {journal} {\bibinfo  {journal} {arXiv preprint
  arXiv:2306.16227}\ } (\bibinfo {year} {2023}{\natexlab{d}})},\ \Eprint
  {http://arxiv.org/abs/2306.16227} {arXiv:2306.16227 [astro-ph.CO]}
  \BibitemShut {NoStop}%
\bibitem [{\citenamefont {Smarra}\ \emph {et~al.}(2023)\citenamefont {Smarra}
  \emph {et~al.}}]{EuropeanPulsarTimingArray:2023qbc}%
  \BibitemOpen
  \bibfield  {author} {\bibinfo {author} {\bibfnamefont {Clemente}\
  \bibnamefont {Smarra}} \emph {et~al.} (\bibinfo {collaboration} {European
  Pulsar Timing Array}),\ }\bibfield  {title} {\enquote {\bibinfo {title} {{The
  second data release from the European Pulsar Timing Array: VI. Challenging
  the ultralight dark matter paradigm}},}\ }\href@noop {} {\bibfield  {journal}
  {\bibinfo  {journal} {.}\ } (\bibinfo {year} {2023})},\ \Eprint
  {http://arxiv.org/abs/2306.16228} {arXiv:2306.16228 [astro-ph.HE]}
  \BibitemShut {NoStop}%
\bibitem [{\citenamefont {Reardon}\ \emph
  {et~al.}(2023{\natexlab{b}})\citenamefont {Reardon} \emph
  {et~al.}}]{Reardon:2023zen}%
  \BibitemOpen
  \bibfield  {author} {\bibinfo {author} {\bibfnamefont {Daniel~J.}\
  \bibnamefont {Reardon}} \emph {et~al.},\ }\bibfield  {title} {\enquote
  {\bibinfo {title} {{The Gravitational-wave Background Null Hypothesis:
  Characterizing Noise in Millisecond Pulsar Arrival Times with the Parkes
  Pulsar Timing Array}},}\ }\href {\doibase 10.3847/2041-8213/acdd03}
  {\bibfield  {journal} {\bibinfo  {journal} {Astrophys. J. Lett.}\ }\textbf
  {\bibinfo {volume} {951}},\ \bibinfo {pages} {L7} (\bibinfo {year}
  {2023}{\natexlab{b}})},\ \Eprint {http://arxiv.org/abs/2306.16229}
  {arXiv:2306.16229 [astro-ph.HE]} \BibitemShut {NoStop}%
\bibitem [{\citenamefont {Zic}\ \emph {et~al.}(2023)\citenamefont {Zic} \emph
  {et~al.}}]{Zic:2023gta}%
  \BibitemOpen
  \bibfield  {author} {\bibinfo {author} {\bibfnamefont {Andrew}\ \bibnamefont
  {Zic}} \emph {et~al.},\ }\bibfield  {title} {\enquote {\bibinfo {title} {{The
  Parkes Pulsar Timing Array Third Data Release}},}\ }\href@noop {} {\bibfield
  {journal} {\bibinfo  {journal} {.}\ } (\bibinfo {year} {2023})},\ \Eprint
  {http://arxiv.org/abs/2306.16230} {arXiv:2306.16230 [astro-ph.HE]}
  \BibitemShut {NoStop}%
\bibitem [{\citenamefont {Vagnozzi}(2020)}]{10.1093/mnrasl/slaa203}%
  \BibitemOpen
  \bibfield  {author} {\bibinfo {author} {\bibfnamefont {Sunny}\ \bibnamefont
  {Vagnozzi}},\ }\bibfield  {title} {\enquote {\bibinfo {title} {{Implications
  of the NANOGrav results for inflation}},}\ }\href {\doibase
  10.1093/mnrasl/slaa203} {\bibfield  {journal} {\bibinfo  {journal} {Monthly
  Notices of the Royal Astronomical Society: Letters}\ }\textbf {\bibinfo
  {volume} {502}},\ \bibinfo {pages} {L11--L15} (\bibinfo {year} {2020})},\
  \Eprint
  {http://arxiv.org/abs/https://academic.oup.com/mnrasl/article-pdf/502/1/L11/35873635/slaa203.pdf}
  {https://academic.oup.com/mnrasl/article-pdf/502/1/L11/35873635/slaa203.pdf}
  \BibitemShut {NoStop}%
\bibitem [{\citenamefont {Vagnozzi}(2023)}]{Vagnozzi:2023lwo}%
  \BibitemOpen
  \bibfield  {author} {\bibinfo {author} {\bibfnamefont {Sunny}\ \bibnamefont
  {Vagnozzi}},\ }\bibfield  {title} {\enquote {\bibinfo {title} {{Inflationary
  interpretation of the stochastic gravitational wave background signal
  detected by pulsar timing array experiments}},}\ }\href@noop {} {\bibfield
  {journal} {\bibinfo  {journal} {.}\ } (\bibinfo {year} {2023})},\ \Eprint
  {http://arxiv.org/abs/2306.16912} {arXiv:2306.16912 [astro-ph.CO]}
  \BibitemShut {NoStop}%
\bibitem [{\citenamefont {Datta}(2023)}]{Datta:2023vbs}%
  \BibitemOpen
  \bibfield  {author} {\bibinfo {author} {\bibfnamefont {Satyabrata}\
  \bibnamefont {Datta}},\ }\bibfield  {title} {\enquote {\bibinfo {title}
  {{Inflationary gravitational waves, pulsar timing data and
  low-scale-leptogenesis}},}\ }\href@noop {} {\bibfield  {journal} {\bibinfo
  {journal} {.}\ } (\bibinfo {year} {2023})},\ \Eprint
  {http://arxiv.org/abs/2307.00646} {arXiv:2307.00646 [hep-ph]} \BibitemShut
  {NoStop}%
\bibitem [{\citenamefont {Lazarides}\ \emph {et~al.}(2023)\citenamefont
  {Lazarides}, \citenamefont {Maji},\ and\ \citenamefont
  {Shafi}}]{Lazarides:2023ksx}%
  \BibitemOpen
  \bibfield  {author} {\bibinfo {author} {\bibfnamefont {George}\ \bibnamefont
  {Lazarides}}, \bibinfo {author} {\bibfnamefont {Rinku}\ \bibnamefont {Maji}},
  \ and\ \bibinfo {author} {\bibfnamefont {Qaisar}\ \bibnamefont {Shafi}},\
  }\bibfield  {title} {\enquote {\bibinfo {title} {{Superheavy quasi-stable
  strings and walls bounded by strings in the light of NANOGrav 15 year
  data}},}\ }\href@noop {} {\bibfield  {journal} {\bibinfo  {journal} {.}\ }
  (\bibinfo {year} {2023})},\ \Eprint {http://arxiv.org/abs/2306.17788}
  {arXiv:2306.17788 [hep-ph]} \BibitemShut {NoStop}%
\bibitem [{\citenamefont {Zhao}\ \emph {et~al.}(2023)\citenamefont {Zhao},
  \citenamefont {Zhu}, \citenamefont {Wang},\ and\ \citenamefont
  {Zhang}}]{Zhao:2023joc}%
  \BibitemOpen
  \bibfield  {author} {\bibinfo {author} {\bibfnamefont {Zhi-Chao}\
  \bibnamefont {Zhao}}, \bibinfo {author} {\bibfnamefont {Qing-Hua}\
  \bibnamefont {Zhu}}, \bibinfo {author} {\bibfnamefont {Sai}\ \bibnamefont
  {Wang}}, \ and\ \bibinfo {author} {\bibfnamefont {Xin}\ \bibnamefont
  {Zhang}},\ }\bibfield  {title} {\enquote {\bibinfo {title} {{Exploring the
  Equation of State of the Early Universe: Insights from BBN, CMB, and PTA
  Observations}},}\ }\href@noop {} {\bibfield  {journal} {\bibinfo  {journal}
  {.}\ } (\bibinfo {year} {2023})},\ \Eprint {http://arxiv.org/abs/2307.13574}
  {arXiv:2307.13574 [astro-ph.CO]} \BibitemShut {NoStop}%
\bibitem [{\citenamefont {Das}\ \emph {et~al.}(2023)\citenamefont {Das},
  \citenamefont {Jaman},\ and\ \citenamefont {Sami}}]{Das:2023nmm}%
  \BibitemOpen
  \bibfield  {author} {\bibinfo {author} {\bibfnamefont {Barnali}\ \bibnamefont
  {Das}}, \bibinfo {author} {\bibfnamefont {Nur}\ \bibnamefont {Jaman}}, \ and\
  \bibinfo {author} {\bibfnamefont {M.}~\bibnamefont {Sami}},\ }\bibfield
  {title} {\enquote {\bibinfo {title} {{Gravitational Waves Background
  (NANOGrav) from Quintessential Inflation}},}\ }\href@noop {} {\bibfield
  {journal} {\bibinfo  {journal} {.}\ } (\bibinfo {year} {2023})},\ \Eprint
  {http://arxiv.org/abs/2307.12913} {arXiv:2307.12913 [gr-qc]} \BibitemShut
  {NoStop}%
\bibitem [{\citenamefont {Gorji}\ \emph {et~al.}(2023)\citenamefont {Gorji},
  \citenamefont {Sasaki},\ and\ \citenamefont {Suyama}}]{Gorji:2023sil}%
  \BibitemOpen
  \bibfield  {author} {\bibinfo {author} {\bibfnamefont {Mohammad~Ali}\
  \bibnamefont {Gorji}}, \bibinfo {author} {\bibfnamefont {Misao}\ \bibnamefont
  {Sasaki}}, \ and\ \bibinfo {author} {\bibfnamefont {Teruaki}\ \bibnamefont
  {Suyama}},\ }\bibfield  {title} {\enquote {\bibinfo {title}
  {{Extra-tensor-induced origin for the PTA signal: No primordial black hole
  production}},}\ }\href@noop {} {\bibfield  {journal} {\bibinfo  {journal}
  {.}\ } (\bibinfo {year} {2023})},\ \Eprint {http://arxiv.org/abs/2307.13109}
  {arXiv:2307.13109 [astro-ph.CO]} \BibitemShut {NoStop}%
\bibitem [{\citenamefont {Ahmadvand}\ \emph {et~al.}(2023)\citenamefont
  {Ahmadvand}, \citenamefont {Bian},\ and\ \citenamefont
  {Shakeri}}]{Ahmadvand:2023lpp}%
  \BibitemOpen
  \bibfield  {author} {\bibinfo {author} {\bibfnamefont {Moslem}\ \bibnamefont
  {Ahmadvand}}, \bibinfo {author} {\bibfnamefont {Ligong}\ \bibnamefont
  {Bian}}, \ and\ \bibinfo {author} {\bibfnamefont {Soroush}\ \bibnamefont
  {Shakeri}},\ }\bibfield  {title} {\enquote {\bibinfo {title} {{A Heavy QCD
  Axion model in Light of Pulsar Timing Arrays}},}\ }\href@noop {} {\bibfield
  {journal} {\bibinfo  {journal} {.}\ } (\bibinfo {year} {2023})},\ \Eprint
  {http://arxiv.org/abs/2307.12385} {arXiv:2307.12385 [hep-ph]} \BibitemShut
  {NoStop}%
\bibitem [{\citenamefont {Zhang}\ \emph {et~al.}(2023)\citenamefont {Zhang},
  \citenamefont {Cai}, \citenamefont {Su}, \citenamefont {Wang}, \citenamefont
  {Yu},\ and\ \citenamefont {Zhang}}]{Zhang:2023nrs}%
  \BibitemOpen
  \bibfield  {author} {\bibinfo {author} {\bibfnamefont {Zhao}\ \bibnamefont
  {Zhang}}, \bibinfo {author} {\bibfnamefont {Chengfeng}\ \bibnamefont {Cai}},
  \bibinfo {author} {\bibfnamefont {Yu-Hang}\ \bibnamefont {Su}}, \bibinfo
  {author} {\bibfnamefont {Shiyu}\ \bibnamefont {Wang}}, \bibinfo {author}
  {\bibfnamefont {Zhao-Huan}\ \bibnamefont {Yu}}, \ and\ \bibinfo {author}
  {\bibfnamefont {Hong-Hao}\ \bibnamefont {Zhang}},\ }\bibfield  {title}
  {\enquote {\bibinfo {title} {{Nano-Hertz gravitational waves from collapsing
  domain walls associated with freeze-in dark matter in light of pulsar timing
  array observations}},}\ }\href@noop {} {\bibfield  {journal} {\bibinfo
  {journal} {.}\ } (\bibinfo {year} {2023})},\ \Eprint
  {http://arxiv.org/abs/2307.11495} {arXiv:2307.11495 [hep-ph]} \BibitemShut
  {NoStop}%
\bibitem [{\citenamefont {Bousder}\ \emph {et~al.}(2023)\citenamefont
  {Bousder}, \citenamefont {Riadsolh}, \citenamefont {Fatimy}, \citenamefont
  {Belkacemi},\ and\ \citenamefont {Ez-Zahraouy}}]{Bousder:2023ida}%
  \BibitemOpen
  \bibfield  {author} {\bibinfo {author} {\bibfnamefont {M.}~\bibnamefont
  {Bousder}}, \bibinfo {author} {\bibfnamefont {A.}~\bibnamefont {Riadsolh}},
  \bibinfo {author} {\bibfnamefont {A.~El}\ \bibnamefont {Fatimy}}, \bibinfo
  {author} {\bibfnamefont {M.~El}\ \bibnamefont {Belkacemi}}, \ and\ \bibinfo
  {author} {\bibfnamefont {H.}~\bibnamefont {Ez-Zahraouy}},\ }\bibfield
  {title} {\enquote {\bibinfo {title} {{Implications of the NANOGrav results
  for primordial black holes and Hubble tension}},}\ }\href@noop {} {\bibfield
  {journal} {\bibinfo  {journal} {.}\ } (\bibinfo {year} {2023})},\ \Eprint
  {http://arxiv.org/abs/2307.10940} {arXiv:2307.10940 [gr-qc]} \BibitemShut
  {NoStop}%
\bibitem [{\citenamefont {Valbusa~Dall'Armi}\ \emph {et~al.}(2023)\citenamefont
  {Valbusa~Dall'Armi}, \citenamefont {Mierna}, \citenamefont {Matarrese},\ and\
  \citenamefont {Ricciardone}}]{ValbusaDallArmi:2023nqn}%
  \BibitemOpen
  \bibfield  {author} {\bibinfo {author} {\bibfnamefont {Lorenzo}\ \bibnamefont
  {Valbusa~Dall'Armi}}, \bibinfo {author} {\bibfnamefont {Alina}\ \bibnamefont
  {Mierna}}, \bibinfo {author} {\bibfnamefont {Sabino}\ \bibnamefont
  {Matarrese}}, \ and\ \bibinfo {author} {\bibfnamefont {Angelo}\ \bibnamefont
  {Ricciardone}},\ }\bibfield  {title} {\enquote {\bibinfo {title} {{Adiabatic
  or Non-Adiabatic? Unraveling the Nature of Initial Conditions in the
  Cosmological Gravitational Wave Background}},}\ }\href@noop {} {\bibfield
  {journal} {\bibinfo  {journal} {.}\ } (\bibinfo {year} {2023})},\ \Eprint
  {http://arxiv.org/abs/2307.11043} {arXiv:2307.11043 [astro-ph.CO]}
  \BibitemShut {NoStop}%
\bibitem [{\citenamefont {Cui}\ \emph {et~al.}(2023)\citenamefont {Cui},
  \citenamefont {Kumar}, \citenamefont {Sundrum},\ and\ \citenamefont
  {Tsai}}]{Cui:2023dlo}%
  \BibitemOpen
  \bibfield  {author} {\bibinfo {author} {\bibfnamefont {Yanou}\ \bibnamefont
  {Cui}}, \bibinfo {author} {\bibfnamefont {Soubhik}\ \bibnamefont {Kumar}},
  \bibinfo {author} {\bibfnamefont {Raman}\ \bibnamefont {Sundrum}}, \ and\
  \bibinfo {author} {\bibfnamefont {Yuhsin}\ \bibnamefont {Tsai}},\ }\bibfield
  {title} {\enquote {\bibinfo {title} {{Unraveling Cosmological Anisotropies
  within Stochastic Gravitational Wave Backgrounds}},}\ }\href@noop {}
  {\bibfield  {journal} {\bibinfo  {journal} {.}\ } (\bibinfo {year} {2023})},\
  \Eprint {http://arxiv.org/abs/2307.10360} {arXiv:2307.10360 [astro-ph.CO]}
  \BibitemShut {NoStop}%
\bibitem [{\citenamefont {Basilakos}\ \emph {et~al.}(2023)\citenamefont
  {Basilakos}, \citenamefont {Nanopoulos}, \citenamefont {Papanikolaou},
  \citenamefont {Saridakis},\ and\ \citenamefont
  {Tzerefos}}]{Basilakos:2023xof}%
  \BibitemOpen
  \bibfield  {author} {\bibinfo {author} {\bibfnamefont {Spyros}\ \bibnamefont
  {Basilakos}}, \bibinfo {author} {\bibfnamefont {Dimitri~V.}\ \bibnamefont
  {Nanopoulos}}, \bibinfo {author} {\bibfnamefont {Theodoros}\ \bibnamefont
  {Papanikolaou}}, \bibinfo {author} {\bibfnamefont {Emmanuel~N.}\ \bibnamefont
  {Saridakis}}, \ and\ \bibinfo {author} {\bibfnamefont {Charalampos}\
  \bibnamefont {Tzerefos}},\ }\bibfield  {title} {\enquote {\bibinfo {title}
  {{Gravitational wave signatures of no-scale Supergravity in NANOGrav and
  beyond}},}\ }\href@noop {} {\bibfield  {journal} {\bibinfo  {journal} {.}\ }
  (\bibinfo {year} {2023})},\ \Eprint {http://arxiv.org/abs/2307.08601}
  {arXiv:2307.08601 [hep-th]} \BibitemShut {NoStop}%
\bibitem [{\citenamefont {Balaji}\ \emph {et~al.}(2023)\citenamefont {Balaji},
  \citenamefont {Dom\`enech},\ and\ \citenamefont
  {Franciolini}}]{Balaji:2023ehk}%
  \BibitemOpen
  \bibfield  {author} {\bibinfo {author} {\bibfnamefont {Shyam}\ \bibnamefont
  {Balaji}}, \bibinfo {author} {\bibfnamefont {Guillem}\ \bibnamefont
  {Dom\`enech}}, \ and\ \bibinfo {author} {\bibfnamefont {Gabriele}\
  \bibnamefont {Franciolini}},\ }\bibfield  {title} {\enquote {\bibinfo {title}
  {{Scalar-induced gravitational wave interpretation of PTA data: the role of
  scalar fluctuation propagation speed}},}\ }\href@noop {} {\bibfield
  {journal} {\bibinfo  {journal} {.}\ } (\bibinfo {year} {2023})},\ \Eprint
  {http://arxiv.org/abs/2307.08552} {arXiv:2307.08552 [gr-qc]} \BibitemShut
  {NoStop}%
\bibitem [{\citenamefont {Gelmini}\ and\ \citenamefont
  {Hyman}(2023)}]{Gelmini:2023kvo}%
  \BibitemOpen
  \bibfield  {author} {\bibinfo {author} {\bibfnamefont {Graciela~B.}\
  \bibnamefont {Gelmini}}\ and\ \bibinfo {author} {\bibfnamefont {Jonah}\
  \bibnamefont {Hyman}},\ }\bibfield  {title} {\enquote {\bibinfo {title}
  {{Catastrogenesis with unstable ALPs as the origin of the NANOGrav 15 yr
  gravitational wave signal}},}\ }\href@noop {} {\bibfield  {journal} {\bibinfo
   {journal} {.}\ } (\bibinfo {year} {2023})},\ \Eprint
  {http://arxiv.org/abs/2307.07665} {arXiv:2307.07665 [hep-ph]} \BibitemShut
  {NoStop}%
\bibitem [{\citenamefont {Yamada}\ and\ \citenamefont
  {Yonekura}(2023)}]{Yamada:2023thl}%
  \BibitemOpen
  \bibfield  {author} {\bibinfo {author} {\bibfnamefont {Masaki}\ \bibnamefont
  {Yamada}}\ and\ \bibinfo {author} {\bibfnamefont {Kazuya}\ \bibnamefont
  {Yonekura}},\ }\bibfield  {title} {\enquote {\bibinfo {title} {{Dark baryon
  from pure Yang-Mills theory and its GW signature from cosmic strings}},}\
  }\href {\doibase 10.1007/JHEP09(2023)197} {\bibfield  {journal} {\bibinfo
  {journal} {JHEP}\ }\textbf {\bibinfo {volume} {09}},\ \bibinfo {pages} {197}
  (\bibinfo {year} {2023})},\ \Eprint {http://arxiv.org/abs/2307.06586}
  {arXiv:2307.06586 [hep-ph]} \BibitemShut {NoStop}%
\bibitem [{\citenamefont {Babichev}\ \emph {et~al.}(2023)\citenamefont
  {Babichev}, \citenamefont {Gorbunov}, \citenamefont {Ramazanov},
  \citenamefont {Samanta},\ and\ \citenamefont {Vikman}}]{Babichev:2023pbf}%
  \BibitemOpen
  \bibfield  {author} {\bibinfo {author} {\bibfnamefont {E.}~\bibnamefont
  {Babichev}}, \bibinfo {author} {\bibfnamefont {D.}~\bibnamefont {Gorbunov}},
  \bibinfo {author} {\bibfnamefont {S.}~\bibnamefont {Ramazanov}}, \bibinfo
  {author} {\bibfnamefont {R.}~\bibnamefont {Samanta}}, \ and\ \bibinfo
  {author} {\bibfnamefont {A.}~\bibnamefont {Vikman}},\ }\bibfield  {title}
  {\enquote {\bibinfo {title} {{NANOGrav spectral index $\gamma=3$ from melting
  domain walls}},}\ }\href@noop {} {\bibfield  {journal} {\bibinfo  {journal}
  {.}\ } (\bibinfo {year} {2023})},\ \Eprint {http://arxiv.org/abs/2307.04582}
  {arXiv:2307.04582 [hep-ph]} \BibitemShut {NoStop}%
\bibitem [{\citenamefont {Buchmuller}\ \emph {et~al.}(2023)\citenamefont
  {Buchmuller}, \citenamefont {Domcke},\ and\ \citenamefont
  {Schmitz}}]{Buchmuller:2023aus}%
  \BibitemOpen
  \bibfield  {author} {\bibinfo {author} {\bibfnamefont {Wilfried}\
  \bibnamefont {Buchmuller}}, \bibinfo {author} {\bibfnamefont {Valerie}\
  \bibnamefont {Domcke}}, \ and\ \bibinfo {author} {\bibfnamefont {Kai}\
  \bibnamefont {Schmitz}},\ }\bibfield  {title} {\enquote {\bibinfo {title}
  {{Metastable cosmic strings}},}\ }\href@noop {} {\bibfield  {journal}
  {\bibinfo  {journal} {.}\ } (\bibinfo {year} {2023})},\ \Eprint
  {http://arxiv.org/abs/2307.04691} {arXiv:2307.04691 [hep-ph]} \BibitemShut
  {NoStop}%
\bibitem [{\citenamefont {You}\ \emph {et~al.}(2023)\citenamefont {You},
  \citenamefont {Yi},\ and\ \citenamefont {Wu}}]{You:2023rmn}%
  \BibitemOpen
  \bibfield  {author} {\bibinfo {author} {\bibfnamefont {Zhi-Qiang}\
  \bibnamefont {You}}, \bibinfo {author} {\bibfnamefont {Zhu}\ \bibnamefont
  {Yi}}, \ and\ \bibinfo {author} {\bibfnamefont {You}\ \bibnamefont {Wu}},\
  }\bibfield  {title} {\enquote {\bibinfo {title} {{Constraints on primordial
  curvature power spectrum with pulsar timing arrays}},}\ }\href@noop {}
  {\bibfield  {journal} {\bibinfo  {journal} {.}\ } (\bibinfo {year} {2023})},\
  \Eprint {http://arxiv.org/abs/2307.04419} {arXiv:2307.04419 [gr-qc]}
  \BibitemShut {NoStop}%
\bibitem [{\citenamefont {Salvio}(2023)}]{Salvio:2023ynn}%
  \BibitemOpen
  \bibfield  {author} {\bibinfo {author} {\bibfnamefont {Alberto}\ \bibnamefont
  {Salvio}},\ }\bibfield  {title} {\enquote {\bibinfo {title} {{Supercooling in
  Radiative Symmetry Breaking: Theory Extensions, Gravitational Wave Detection
  and Primordial Black Holes}},}\ }\href@noop {} {\bibfield  {journal}
  {\bibinfo  {journal} {.}\ } (\bibinfo {year} {2023})},\ \Eprint
  {http://arxiv.org/abs/2307.04694} {arXiv:2307.04694 [hep-ph]} \BibitemShut
  {NoStop}%
\bibitem [{\citenamefont {Gouttenoire}(2023)}]{Gouttenoire:2023bqy}%
  \BibitemOpen
  \bibfield  {author} {\bibinfo {author} {\bibfnamefont {Yann}\ \bibnamefont
  {Gouttenoire}},\ }\bibfield  {title} {\enquote {\bibinfo {title}
  {{First-order Phase Transition interpretation of PTA signal produces
  solar-mass Black Holes}},}\ }\href@noop {} {\bibfield  {journal} {\bibinfo
  {journal} {.}\ } (\bibinfo {year} {2023})},\ \Eprint
  {http://arxiv.org/abs/2307.04239} {arXiv:2307.04239 [hep-ph]} \BibitemShut
  {NoStop}%
\bibitem [{\citenamefont {Geller}\ \emph {et~al.}(2023)\citenamefont {Geller},
  \citenamefont {Ghosh}, \citenamefont {Lu},\ and\ \citenamefont
  {Tsai}}]{Geller:2023shn}%
  \BibitemOpen
  \bibfield  {author} {\bibinfo {author} {\bibfnamefont {Michael}\ \bibnamefont
  {Geller}}, \bibinfo {author} {\bibfnamefont {Subhajit}\ \bibnamefont
  {Ghosh}}, \bibinfo {author} {\bibfnamefont {Sida}\ \bibnamefont {Lu}}, \ and\
  \bibinfo {author} {\bibfnamefont {Yuhsin}\ \bibnamefont {Tsai}},\ }\bibfield
  {title} {\enquote {\bibinfo {title} {{Challenges in Interpreting the NANOGrav
  15-Year Data Set as Early Universe Gravitational Waves Produced by ALP
  Induced Instability}},}\ }\href@noop {} {\bibfield  {journal} {\bibinfo
  {journal} {.}\ } (\bibinfo {year} {2023})},\ \Eprint
  {http://arxiv.org/abs/2307.03724} {arXiv:2307.03724 [hep-ph]} \BibitemShut
  {NoStop}%
\bibitem [{\citenamefont {Du}\ \emph {et~al.}(2023)\citenamefont {Du},
  \citenamefont {Huang}, \citenamefont {Wang},\ and\ \citenamefont
  {Zhang}}]{Du:2023qvj}%
  \BibitemOpen
  \bibfield  {author} {\bibinfo {author} {\bibfnamefont {Xiao~Kang}\
  \bibnamefont {Du}}, \bibinfo {author} {\bibfnamefont {Ming~Xia}\ \bibnamefont
  {Huang}}, \bibinfo {author} {\bibfnamefont {Fei}\ \bibnamefont {Wang}}, \
  and\ \bibinfo {author} {\bibfnamefont {Ying~Kai}\ \bibnamefont {Zhang}},\
  }\bibfield  {title} {\enquote {\bibinfo {title} {{Did the nHZ Gravitational
  Waves Signatures Observed By NANOGrav Indicate Multiple Sector SUSY
  Breaking?}}}\ }\href@noop {} {\bibfield  {journal} {\bibinfo  {journal} {.}\
  } (\bibinfo {year} {2023})},\ \Eprint {http://arxiv.org/abs/2307.02938}
  {arXiv:2307.02938 [hep-ph]} \BibitemShut {NoStop}%
\bibitem [{\citenamefont {Servant}\ and\ \citenamefont
  {Simakachorn}(2023)}]{Servant:2023mwt}%
  \BibitemOpen
  \bibfield  {author} {\bibinfo {author} {\bibfnamefont {G\'eraldine}\
  \bibnamefont {Servant}}\ and\ \bibinfo {author} {\bibfnamefont {Peera}\
  \bibnamefont {Simakachorn}},\ }\bibfield  {title} {\enquote {\bibinfo {title}
  {{Constraining Post-Inflationary Axions with Pulsar Timing Arrays}},}\
  }\href@noop {} {\bibfield  {journal} {\bibinfo  {journal} {.}\ } (\bibinfo
  {year} {2023})},\ \Eprint {http://arxiv.org/abs/2307.03121} {arXiv:2307.03121
  [hep-ph]} \BibitemShut {NoStop}%
\bibitem [{\citenamefont {Wu}\ \emph {et~al.}(2023)\citenamefont {Wu},
  \citenamefont {Chen},\ and\ \citenamefont {Huang}}]{Wu:2023hsa}%
  \BibitemOpen
  \bibfield  {author} {\bibinfo {author} {\bibfnamefont {Yu-Mei}\ \bibnamefont
  {Wu}}, \bibinfo {author} {\bibfnamefont {Zu-Cheng}\ \bibnamefont {Chen}}, \
  and\ \bibinfo {author} {\bibfnamefont {Qing-Guo}\ \bibnamefont {Huang}},\
  }\bibfield  {title} {\enquote {\bibinfo {title} {{Cosmological Interpretation
  for the Stochastic Signal in Pulsar Timing Arrays}},}\ }\href@noop {}
  {\bibfield  {journal} {\bibinfo  {journal} {.}\ } (\bibinfo {year} {2023})},\
  \Eprint {http://arxiv.org/abs/2307.03141} {arXiv:2307.03141 [astro-ph.CO]}
  \BibitemShut {NoStop}%
\bibitem [{\citenamefont {Li}(2023)}]{Li:2023tdx}%
  \BibitemOpen
  \bibfield  {author} {\bibinfo {author} {\bibfnamefont {Xiu-Fei}\ \bibnamefont
  {Li}},\ }\bibfield  {title} {\enquote {\bibinfo {title} {{Probing the high
  temperature symmetry breaking with gravitational waves from domain walls}},}\
  }\href@noop {} {\bibfield  {journal} {\bibinfo  {journal} {.}\ } (\bibinfo
  {year} {2023})},\ \Eprint {http://arxiv.org/abs/2307.03163} {arXiv:2307.03163
  [hep-ph]} \BibitemShut {NoStop}%
\bibitem [{\citenamefont {Liu}\ \emph {et~al.}(2023{\natexlab{a}})\citenamefont
  {Liu}, \citenamefont {Chen},\ and\ \citenamefont {Huang}}]{Liu:2023pau}%
  \BibitemOpen
  \bibfield  {author} {\bibinfo {author} {\bibfnamefont {Lang}\ \bibnamefont
  {Liu}}, \bibinfo {author} {\bibfnamefont {Zu-Cheng}\ \bibnamefont {Chen}}, \
  and\ \bibinfo {author} {\bibfnamefont {Qing-Guo}\ \bibnamefont {Huang}},\
  }\bibfield  {title} {\enquote {\bibinfo {title} {{Probing the equation of
  state of the early Universe with pulsar timing arrays}},}\ }\href@noop {}
  {\bibfield  {journal} {\bibinfo  {journal} {.}\ } (\bibinfo {year}
  {2023}{\natexlab{a}})},\ \Eprint {http://arxiv.org/abs/2307.14911}
  {arXiv:2307.14911 [astro-ph.CO]} \BibitemShut {NoStop}%
\bibitem [{\citenamefont {Liu}\ \emph {et~al.}(2023{\natexlab{b}})\citenamefont
  {Liu}, \citenamefont {Chen},\ and\ \citenamefont {Huang}}]{Liu:2023ymk}%
  \BibitemOpen
  \bibfield  {author} {\bibinfo {author} {\bibfnamefont {Lang}\ \bibnamefont
  {Liu}}, \bibinfo {author} {\bibfnamefont {Zu-Cheng}\ \bibnamefont {Chen}}, \
  and\ \bibinfo {author} {\bibfnamefont {Qing-Guo}\ \bibnamefont {Huang}},\
  }\bibfield  {title} {\enquote {\bibinfo {title} {{Implications for the
  non-Gaussianity of curvature perturbation from pulsar timing arrays}},}\
  }\href@noop {} {\bibfield  {journal} {\bibinfo  {journal} {.}\ } (\bibinfo
  {year} {2023}{\natexlab{b}})},\ \Eprint {http://arxiv.org/abs/2307.01102}
  {arXiv:2307.01102 [astro-ph.CO]} \BibitemShut {NoStop}%
\bibitem [{\citenamefont {Oikonomou}(2023)}]{Oikonomou:2023qfz}%
  \BibitemOpen
  \bibfield  {author} {\bibinfo {author} {\bibfnamefont {V.~K.}\ \bibnamefont
  {Oikonomou}},\ }\bibfield  {title} {\enquote {\bibinfo {title} {{Flat energy
  spectrum of primordial gravitational waves versus peaks and the NANOGrav 2023
  observation}},}\ }\href {\doibase 10.1103/PhysRevD.108.043516} {\bibfield
  {journal} {\bibinfo  {journal} {Phys. Rev. D}\ }\textbf {\bibinfo {volume}
  {108}},\ \bibinfo {pages} {043516} (\bibinfo {year} {2023})},\ \Eprint
  {http://arxiv.org/abs/2306.17351} {arXiv:2306.17351 [astro-ph.CO]}
  \BibitemShut {NoStop}%
\bibitem [{\citenamefont {Figueroa}\ \emph {et~al.}(2023)\citenamefont
  {Figueroa}, \citenamefont {Pieroni}, \citenamefont {Ricciardone},\ and\
  \citenamefont {Simakachorn}}]{Figueroa:2023zhu}%
  \BibitemOpen
  \bibfield  {author} {\bibinfo {author} {\bibfnamefont {Daniel~G.}\
  \bibnamefont {Figueroa}}, \bibinfo {author} {\bibfnamefont {Mauro}\
  \bibnamefont {Pieroni}}, \bibinfo {author} {\bibfnamefont {Angelo}\
  \bibnamefont {Ricciardone}}, \ and\ \bibinfo {author} {\bibfnamefont {Peera}\
  \bibnamefont {Simakachorn}},\ }\bibfield  {title} {\enquote {\bibinfo {title}
  {{Cosmological Background Interpretation of Pulsar Timing Array Data}},}\
  }\href@noop {} {\bibfield  {journal} {\bibinfo  {journal} {.}\ } (\bibinfo
  {year} {2023})},\ \Eprint {http://arxiv.org/abs/2307.02399} {arXiv:2307.02399
  [astro-ph.CO]} \BibitemShut {NoStop}%
\bibitem [{\citenamefont {Unal}\ \emph {et~al.}(2023)\citenamefont {Unal},
  \citenamefont {Papageorgiou},\ and\ \citenamefont {Obata}}]{Unal:2023srk}%
  \BibitemOpen
  \bibfield  {author} {\bibinfo {author} {\bibfnamefont {Caner}\ \bibnamefont
  {Unal}}, \bibinfo {author} {\bibfnamefont {Alexandros}\ \bibnamefont
  {Papageorgiou}}, \ and\ \bibinfo {author} {\bibfnamefont {Ippei}\
  \bibnamefont {Obata}},\ }\bibfield  {title} {\enquote {\bibinfo {title}
  {{Axion-Gauge Dynamics During Inflation as the Origin of Pulsar Timing Array
  Signals and Primordial Black Holes}},}\ }\href@noop {} {\bibfield  {journal}
  {\bibinfo  {journal} {.}\ } (\bibinfo {year} {2023})},\ \Eprint
  {http://arxiv.org/abs/2307.02322} {arXiv:2307.02322 [astro-ph.CO]}
  \BibitemShut {NoStop}%
\bibitem [{\citenamefont {Niu}\ and\ \citenamefont
  {Rahat}(2023)}]{Niu:2023bsr}%
  \BibitemOpen
  \bibfield  {author} {\bibinfo {author} {\bibfnamefont {Xuce}\ \bibnamefont
  {Niu}}\ and\ \bibinfo {author} {\bibfnamefont {Moinul~Hossain}\ \bibnamefont
  {Rahat}},\ }\bibfield  {title} {\enquote {\bibinfo {title} {{NANOGrav signal
  from axion inflation}},}\ }\href@noop {} {\bibfield  {journal} {\bibinfo
  {journal} {.}\ } (\bibinfo {year} {2023})},\ \Eprint
  {http://arxiv.org/abs/2307.01192} {arXiv:2307.01192 [hep-ph]} \BibitemShut
  {NoStop}%
\bibitem [{\citenamefont {Cai}\ \emph {et~al.}(2023)\citenamefont {Cai},
  \citenamefont {He}, \citenamefont {Ma}, \citenamefont {Yan},\ and\
  \citenamefont {Yuan}}]{Cai:2023dls}%
  \BibitemOpen
  \bibfield  {author} {\bibinfo {author} {\bibfnamefont {Yi-Fu}\ \bibnamefont
  {Cai}}, \bibinfo {author} {\bibfnamefont {Xin-Chen}\ \bibnamefont {He}},
  \bibinfo {author} {\bibfnamefont {Xiaohan}\ \bibnamefont {Ma}}, \bibinfo
  {author} {\bibfnamefont {Sheng-Feng}\ \bibnamefont {Yan}}, \ and\ \bibinfo
  {author} {\bibfnamefont {Guan-Wen}\ \bibnamefont {Yuan}},\ }\bibfield
  {title} {\enquote {\bibinfo {title} {{Limits on scalar-induced gravitational
  waves from the stochastic background by pulsar timing array observations}},}\
  }\href@noop {} {\bibfield  {journal} {\bibinfo  {journal} {.}\ } (\bibinfo
  {year} {2023})},\ \Eprint {http://arxiv.org/abs/2306.17822} {arXiv:2306.17822
  [gr-qc]} \BibitemShut {NoStop}%
\bibitem [{\citenamefont {Ben-Dayan}(2016)}]{Ben-Dayan:2016iks}%
  \BibitemOpen
  \bibfield  {author} {\bibinfo {author} {\bibfnamefont {Ido}\ \bibnamefont
  {Ben-Dayan}},\ }\bibfield  {title} {\enquote {\bibinfo {title}
  {{Gravitational Waves in Bouncing Cosmologies from Gauge Field
  Production}},}\ }\href {\doibase 10.1088/1475-7516/2016/09/017} {\bibfield
  {journal} {\bibinfo  {journal} {JCAP}\ }\textbf {\bibinfo {volume} {09}},\
  \bibinfo {pages} {017} (\bibinfo {year} {2016})},\ \Eprint
  {http://arxiv.org/abs/1604.07899} {arXiv:1604.07899 [astro-ph.CO]}
  \BibitemShut {NoStop}%
\bibitem [{\citenamefont {Ben-Dayan}\ and\ \citenamefont
  {Kupferman}(2019)}]{Ben-Dayan:2018ksd}%
  \BibitemOpen
  \bibfield  {author} {\bibinfo {author} {\bibfnamefont {Ido}\ \bibnamefont
  {Ben-Dayan}}\ and\ \bibinfo {author} {\bibfnamefont {Judy}\ \bibnamefont
  {Kupferman}},\ }\bibfield  {title} {\enquote {\bibinfo {title} {{Sourced
  scalar fluctuations in bouncing cosmology}},}\ }\href {\doibase
  10.1088/1475-7516/2019/07/050} {\bibfield  {journal} {\bibinfo  {journal}
  {JCAP}\ }\textbf {\bibinfo {volume} {07}},\ \bibinfo {pages} {050} (\bibinfo
  {year} {2019})},\ \bibinfo {note} {[Erratum: JCAP 12, E01 (2020)]},\ \Eprint
  {http://arxiv.org/abs/1812.06970} {arXiv:1812.06970 [gr-qc]} \BibitemShut
  {NoStop}%
\bibitem [{\citenamefont {Martin}\ \emph {et~al.}(2015)\citenamefont {Martin},
  \citenamefont {Ringeval},\ and\ \citenamefont {Vennin}}]{Martin:2014nya}%
  \BibitemOpen
  \bibfield  {author} {\bibinfo {author} {\bibfnamefont {Jerome}\ \bibnamefont
  {Martin}}, \bibinfo {author} {\bibfnamefont {Christophe}\ \bibnamefont
  {Ringeval}}, \ and\ \bibinfo {author} {\bibfnamefont {Vincent}\ \bibnamefont
  {Vennin}},\ }\bibfield  {title} {\enquote {\bibinfo {title} {{Observing
  Inflationary Reheating}},}\ }\href {\doibase 10.1103/PhysRevLett.114.081303}
  {\bibfield  {journal} {\bibinfo  {journal} {Phys. Rev. Lett.}\ }\textbf
  {\bibinfo {volume} {114}},\ \bibinfo {pages} {081303} (\bibinfo {year}
  {2015})},\ \Eprint {http://arxiv.org/abs/1410.7958} {arXiv:1410.7958
  [astro-ph.CO]} \BibitemShut {NoStop}%
\bibitem [{\citenamefont {Dai}\ \emph {et~al.}(2014)\citenamefont {Dai},
  \citenamefont {Kamionkowski},\ and\ \citenamefont {Wang}}]{Dai:2014jja}%
  \BibitemOpen
  \bibfield  {author} {\bibinfo {author} {\bibfnamefont {Liang}\ \bibnamefont
  {Dai}}, \bibinfo {author} {\bibfnamefont {Marc}\ \bibnamefont
  {Kamionkowski}}, \ and\ \bibinfo {author} {\bibfnamefont {Junpu}\
  \bibnamefont {Wang}},\ }\bibfield  {title} {\enquote {\bibinfo {title}
  {{Reheating constraints to inflationary models}},}\ }\href {\doibase
  10.1103/PhysRevLett.113.041302} {\bibfield  {journal} {\bibinfo  {journal}
  {Phys. Rev. Lett.}\ }\textbf {\bibinfo {volume} {113}},\ \bibinfo {pages}
  {041302} (\bibinfo {year} {2014})},\ \Eprint {http://arxiv.org/abs/1404.6704}
  {arXiv:1404.6704 [astro-ph.CO]} \BibitemShut {NoStop}%
\bibitem [{\citenamefont {Lozanov}(2019)}]{Lozanov:2019jxc}%
  \BibitemOpen
  \bibfield  {author} {\bibinfo {author} {\bibfnamefont {Kaloian~D.}\
  \bibnamefont {Lozanov}},\ }\bibfield  {title} {\enquote {\bibinfo {title}
  {{Lectures on Reheating after Inflation}},}\ }\href@noop {} {\bibfield
  {journal} {\bibinfo  {journal} {.}\ } (\bibinfo {year} {2019})},\ \Eprint
  {http://arxiv.org/abs/1907.04402} {arXiv:1907.04402 [astro-ph.CO]}
  \BibitemShut {NoStop}%
\bibitem [{\citenamefont {Boyle}\ and\ \citenamefont
  {Buonanno}(2008)}]{Boyle:2007zx}%
  \BibitemOpen
  \bibfield  {author} {\bibinfo {author} {\bibfnamefont {Latham~A.}\
  \bibnamefont {Boyle}}\ and\ \bibinfo {author} {\bibfnamefont {Alessandra}\
  \bibnamefont {Buonanno}},\ }\bibfield  {title} {\enquote {\bibinfo {title}
  {{Relating gravitational wave constraints from primordial nucleosynthesis,
  pulsar timing, laser interferometers, and the CMB: Implications for the early
  Universe}},}\ }\href {\doibase 10.1103/PhysRevD.78.043531} {\bibfield
  {journal} {\bibinfo  {journal} {Phys. Rev. D}\ }\textbf {\bibinfo {volume}
  {78}},\ \bibinfo {pages} {043531} (\bibinfo {year} {2008})},\ \Eprint
  {http://arxiv.org/abs/0708.2279} {arXiv:0708.2279 [astro-ph]} \BibitemShut
  {NoStop}%
\bibitem [{\citenamefont {Battefeld}\ and\ \citenamefont
  {Peter}(2015)}]{Battefeld:2014uga}%
  \BibitemOpen
  \bibfield  {author} {\bibinfo {author} {\bibfnamefont {D.}~\bibnamefont
  {Battefeld}}\ and\ \bibinfo {author} {\bibfnamefont {Patrick}\ \bibnamefont
  {Peter}},\ }\bibfield  {title} {\enquote {\bibinfo {title} {{A Critical
  Review of Classical Bouncing Cosmologies}},}\ }\href {\doibase
  10.1016/j.physrep.2014.12.004} {\bibfield  {journal} {\bibinfo  {journal}
  {Phys. Rept.}\ }\textbf {\bibinfo {volume} {571}},\ \bibinfo {pages} {1--66}
  (\bibinfo {year} {2015})},\ \Eprint {http://arxiv.org/abs/1406.2790}
  {arXiv:1406.2790 [astro-ph.CO]} \BibitemShut {NoStop}%
\bibitem [{\citenamefont {Khoury}\ \emph {et~al.}(2001)\citenamefont {Khoury},
  \citenamefont {Ovrut}, \citenamefont {Steinhardt},\ and\ \citenamefont
  {Turok}}]{Khoury:2001wf}%
  \BibitemOpen
  \bibfield  {author} {\bibinfo {author} {\bibfnamefont {Justin}\ \bibnamefont
  {Khoury}}, \bibinfo {author} {\bibfnamefont {Burt~A.}\ \bibnamefont {Ovrut}},
  \bibinfo {author} {\bibfnamefont {Paul~J.}\ \bibnamefont {Steinhardt}}, \
  and\ \bibinfo {author} {\bibfnamefont {Neil}\ \bibnamefont {Turok}},\
  }\bibfield  {title} {\enquote {\bibinfo {title} {{The Ekpyrotic universe:
  Colliding branes and the origin of the hot big bang}},}\ }\href {\doibase
  10.1103/PhysRevD.64.123522} {\bibfield  {journal} {\bibinfo  {journal} {Phys.
  Rev. D}\ }\textbf {\bibinfo {volume} {64}},\ \bibinfo {pages} {123522}
  (\bibinfo {year} {2001})},\ \Eprint {http://arxiv.org/abs/hep-th/0103239}
  {arXiv:hep-th/0103239} \BibitemShut {NoStop}%
\bibitem [{\citenamefont {Kobayashi}\ \emph {et~al.}(2010)\citenamefont
  {Kobayashi}, \citenamefont {Yamaguchi},\ and\ \citenamefont
  {Yokoyama}}]{Kobayashi:2010cm}%
  \BibitemOpen
  \bibfield  {author} {\bibinfo {author} {\bibfnamefont {Tsutomu}\ \bibnamefont
  {Kobayashi}}, \bibinfo {author} {\bibfnamefont {Masahide}\ \bibnamefont
  {Yamaguchi}}, \ and\ \bibinfo {author} {\bibfnamefont {Jun'ichi}\
  \bibnamefont {Yokoyama}},\ }\bibfield  {title} {\enquote {\bibinfo {title}
  {{G-inflation: Inflation driven by the Galileon field}},}\ }\href {\doibase
  10.1103/PhysRevLett.105.231302} {\bibfield  {journal} {\bibinfo  {journal}
  {Phys. Rev. Lett.}\ }\textbf {\bibinfo {volume} {105}},\ \bibinfo {pages}
  {231302} (\bibinfo {year} {2010})},\ \Eprint {http://arxiv.org/abs/1008.0603}
  {arXiv:1008.0603 [hep-th]} \BibitemShut {NoStop}%
\bibitem [{\citenamefont {Tahara}\ and\ \citenamefont
  {Kobayashi}(2020)}]{Tahara:2020fmn}%
  \BibitemOpen
  \bibfield  {author} {\bibinfo {author} {\bibfnamefont {Hiroaki W.~H.}\
  \bibnamefont {Tahara}}\ and\ \bibinfo {author} {\bibfnamefont {Tsutomu}\
  \bibnamefont {Kobayashi}},\ }\bibfield  {title} {\enquote {\bibinfo {title}
  {{Nanohertz gravitational waves from a null-energy-condition violation in the
  early universe}},}\ }\href {\doibase 10.1103/PhysRevD.102.123533} {\bibfield
  {journal} {\bibinfo  {journal} {Phys. Rev. D}\ }\textbf {\bibinfo {volume}
  {102}},\ \bibinfo {pages} {123533} (\bibinfo {year} {2020})},\ \Eprint
  {http://arxiv.org/abs/2011.01605} {arXiv:2011.01605 [gr-qc]} \BibitemShut
  {NoStop}%
\bibitem [{\citenamefont {Piao}\ and\ \citenamefont
  {Zhang}(2004)}]{Piao:2004tq}%
  \BibitemOpen
  \bibfield  {author} {\bibinfo {author} {\bibfnamefont {Yun-Song}\
  \bibnamefont {Piao}}\ and\ \bibinfo {author} {\bibfnamefont {Yuan-Zhong}\
  \bibnamefont {Zhang}},\ }\bibfield  {title} {\enquote {\bibinfo {title}
  {{Phantom inflation and primordial perturbation spectrum}},}\ }\href
  {\doibase 10.1103/PhysRevD.70.063513} {\bibfield  {journal} {\bibinfo
  {journal} {Phys. Rev. D}\ }\textbf {\bibinfo {volume} {70}},\ \bibinfo
  {pages} {063513} (\bibinfo {year} {2004})},\ \Eprint
  {http://arxiv.org/abs/astro-ph/0401231} {arXiv:astro-ph/0401231} \BibitemShut
  {NoStop}%
\bibitem [{\citenamefont {Gruzinov}(2004)}]{Gruzinov:2004ty}%
  \BibitemOpen
  \bibfield  {author} {\bibinfo {author} {\bibfnamefont {Andrei}\ \bibnamefont
  {Gruzinov}},\ }\bibfield  {title} {\enquote {\bibinfo {title} {{Elastic
  inflation}},}\ }\href {\doibase 10.1103/PhysRevD.70.063518} {\bibfield
  {journal} {\bibinfo  {journal} {Phys. Rev. D}\ }\textbf {\bibinfo {volume}
  {70}},\ \bibinfo {pages} {063518} (\bibinfo {year} {2004})},\ \Eprint
  {http://arxiv.org/abs/astro-ph/0404548} {arXiv:astro-ph/0404548} \BibitemShut
  {NoStop}%
\bibitem [{\citenamefont {Satoh}\ and\ \citenamefont
  {Soda}(2008)}]{Satoh:2008ck}%
  \BibitemOpen
  \bibfield  {author} {\bibinfo {author} {\bibfnamefont {Masaki}\ \bibnamefont
  {Satoh}}\ and\ \bibinfo {author} {\bibfnamefont {Jiro}\ \bibnamefont
  {Soda}},\ }\bibfield  {title} {\enquote {\bibinfo {title} {{Higher Curvature
  Corrections to Primordial Fluctuations in Slow-roll Inflation}},}\ }\href
  {\doibase 10.1088/1475-7516/2008/09/019} {\bibfield  {journal} {\bibinfo
  {journal} {JCAP}\ }\textbf {\bibinfo {volume} {09}},\ \bibinfo {pages} {019}
  (\bibinfo {year} {2008})},\ \Eprint {http://arxiv.org/abs/0806.4594}
  {arXiv:0806.4594 [astro-ph]} \BibitemShut {NoStop}%
\bibitem [{\citenamefont {Mishima}\ and\ \citenamefont
  {Kobayashi}(2020)}]{Mishima:2019vlh}%
  \BibitemOpen
  \bibfield  {author} {\bibinfo {author} {\bibfnamefont {Yosuke}\ \bibnamefont
  {Mishima}}\ and\ \bibinfo {author} {\bibfnamefont {Tsutomu}\ \bibnamefont
  {Kobayashi}},\ }\bibfield  {title} {\enquote {\bibinfo {title} {{Revisiting
  slow-roll dynamics and the tensor tilt in general single-field inflation}},}\
  }\href {\doibase 10.1103/PhysRevD.101.043536} {\bibfield  {journal} {\bibinfo
   {journal} {Phys. Rev. D}\ }\textbf {\bibinfo {volume} {101}},\ \bibinfo
  {pages} {043536} (\bibinfo {year} {2020})},\ \Eprint
  {http://arxiv.org/abs/1911.02143} {arXiv:1911.02143 [gr-qc]} \BibitemShut
  {NoStop}%
\bibitem [{\citenamefont {Cai}\ \emph {et~al.}(2015)\citenamefont {Cai},
  \citenamefont {Gong}, \citenamefont {Pi}, \citenamefont {Saridakis},\ and\
  \citenamefont {Wu}}]{Cai:2014uka}%
  \BibitemOpen
  \bibfield  {author} {\bibinfo {author} {\bibfnamefont {Yi-Fu}\ \bibnamefont
  {Cai}}, \bibinfo {author} {\bibfnamefont {Jinn-Ouk}\ \bibnamefont {Gong}},
  \bibinfo {author} {\bibfnamefont {Shi}\ \bibnamefont {Pi}}, \bibinfo {author}
  {\bibfnamefont {Emmanuel~N.}\ \bibnamefont {Saridakis}}, \ and\ \bibinfo
  {author} {\bibfnamefont {Shang-Yu}\ \bibnamefont {Wu}},\ }\bibfield  {title}
  {\enquote {\bibinfo {title} {{On the possibility of blue tensor spectrum
  within single field inflation}},}\ }\href {\doibase
  10.1016/j.nuclphysb.2015.09.025} {\bibfield  {journal} {\bibinfo  {journal}
  {Nucl. Phys. B}\ }\textbf {\bibinfo {volume} {900}},\ \bibinfo {pages}
  {517--532} (\bibinfo {year} {2015})},\ \Eprint
  {http://arxiv.org/abs/1412.7241} {arXiv:1412.7241 [hep-th]} \BibitemShut
  {NoStop}%
\bibitem [{\citenamefont {Gong}(2014)}]{Gong:2014qga}%
  \BibitemOpen
  \bibfield  {author} {\bibinfo {author} {\bibfnamefont {Jinn-Ouk}\
  \bibnamefont {Gong}},\ }\bibfield  {title} {\enquote {\bibinfo {title} {{Blue
  running of the primordial tensor spectrum}},}\ }\href {\doibase
  10.1088/1475-7516/2014/07/022} {\bibfield  {journal} {\bibinfo  {journal}
  {JCAP}\ }\textbf {\bibinfo {volume} {07}},\ \bibinfo {pages} {022} (\bibinfo
  {year} {2014})},\ \Eprint {http://arxiv.org/abs/1403.5163} {arXiv:1403.5163
  [astro-ph.CO]} \BibitemShut {NoStop}%
\bibitem [{\citenamefont {Chen}\ \emph {et~al.}(2013)\citenamefont {Chen},
  \citenamefont {Firouzjahi}, \citenamefont {Namjoo},\ and\ \citenamefont
  {Sasaki}}]{Chen:2013aj}%
  \BibitemOpen
  \bibfield  {author} {\bibinfo {author} {\bibfnamefont {Xingang}\ \bibnamefont
  {Chen}}, \bibinfo {author} {\bibfnamefont {Hassan}\ \bibnamefont
  {Firouzjahi}}, \bibinfo {author} {\bibfnamefont {Mohammad~Hossein}\
  \bibnamefont {Namjoo}}, \ and\ \bibinfo {author} {\bibfnamefont {Misao}\
  \bibnamefont {Sasaki}},\ }\bibfield  {title} {\enquote {\bibinfo {title} {{A
  Single Field Inflation Model with Large Local Non-Gaussianity}},}\ }\href
  {\doibase 10.1209/0295-5075/102/59001} {\bibfield  {journal} {\bibinfo
  {journal} {EPL}\ }\textbf {\bibinfo {volume} {102}},\ \bibinfo {pages}
  {59001} (\bibinfo {year} {2013})},\ \Eprint {http://arxiv.org/abs/1301.5699}
  {arXiv:1301.5699 [hep-th]} \BibitemShut {NoStop}%
\bibitem [{\citenamefont {Lin}(2014)}]{Lin:2013sja}%
  \BibitemOpen
  \bibfield  {author} {\bibinfo {author} {\bibfnamefont {Chunshan}\
  \bibnamefont {Lin}},\ }\bibfield  {title} {\enquote {\bibinfo {title}
  {{Massive Graviton on a Spatial Condensate}},}\ }\href {\doibase
  10.1016/j.physletb.2014.09.065} {\bibfield  {journal} {\bibinfo  {journal}
  {Phys. Lett. B}\ }\textbf {\bibinfo {volume} {738}},\ \bibinfo {pages}
  {386--390} (\bibinfo {year} {2014})},\ \Eprint
  {http://arxiv.org/abs/1307.2574} {arXiv:1307.2574 [hep-th]} \BibitemShut
  {NoStop}%
\bibitem [{\citenamefont {Gumrukcuoglu}\ \emph {et~al.}(2012)\citenamefont
  {Gumrukcuoglu}, \citenamefont {Kuroyanagi}, \citenamefont {Lin},
  \citenamefont {Mukohyama},\ and\ \citenamefont
  {Tanahashi}}]{Gumrukcuoglu:2012wt}%
  \BibitemOpen
  \bibfield  {author} {\bibinfo {author} {\bibfnamefont {A.~Emir}\ \bibnamefont
  {Gumrukcuoglu}}, \bibinfo {author} {\bibfnamefont {Sachiko}\ \bibnamefont
  {Kuroyanagi}}, \bibinfo {author} {\bibfnamefont {Chunshan}\ \bibnamefont
  {Lin}}, \bibinfo {author} {\bibfnamefont {Shinji}\ \bibnamefont {Mukohyama}},
  \ and\ \bibinfo {author} {\bibfnamefont {Norihiro}\ \bibnamefont
  {Tanahashi}},\ }\bibfield  {title} {\enquote {\bibinfo {title}
  {{Gravitational wave signal from massive gravity}},}\ }\href {\doibase
  10.1088/0264-9381/29/23/235026} {\bibfield  {journal} {\bibinfo  {journal}
  {Class. Quant. Grav.}\ }\textbf {\bibinfo {volume} {29}},\ \bibinfo {pages}
  {235026} (\bibinfo {year} {2012})},\ \Eprint {http://arxiv.org/abs/1208.5975}
  {arXiv:1208.5975 [hep-th]} \BibitemShut {NoStop}%
\bibitem [{\citenamefont {Barnaby}\ \emph {et~al.}(2012)\citenamefont
  {Barnaby}, \citenamefont {Pajer},\ and\ \citenamefont
  {Peloso}}]{Barnaby:2011qe}%
  \BibitemOpen
  \bibfield  {author} {\bibinfo {author} {\bibfnamefont {Neil}\ \bibnamefont
  {Barnaby}}, \bibinfo {author} {\bibfnamefont {Enrico}\ \bibnamefont {Pajer}},
  \ and\ \bibinfo {author} {\bibfnamefont {Marco}\ \bibnamefont {Peloso}},\
  }\bibfield  {title} {\enquote {\bibinfo {title} {{Gauge Field Production in
  Axion Inflation: Consequences for Monodromy, non-Gaussianity in the CMB, and
  Gravitational Waves at Interferometers}},}\ }\href {\doibase
  10.1103/PhysRevD.85.023525} {\bibfield  {journal} {\bibinfo  {journal} {Phys.
  Rev. D}\ }\textbf {\bibinfo {volume} {85}},\ \bibinfo {pages} {023525}
  (\bibinfo {year} {2012})},\ \Eprint {http://arxiv.org/abs/1110.3327}
  {arXiv:1110.3327 [astro-ph.CO]} \BibitemShut {NoStop}%
\bibitem [{\citenamefont {Wang}\ and\ \citenamefont
  {Xue}(2014)}]{Wang:2014kqa}%
  \BibitemOpen
  \bibfield  {author} {\bibinfo {author} {\bibfnamefont {Yi}~\bibnamefont
  {Wang}}\ and\ \bibinfo {author} {\bibfnamefont {Wei}\ \bibnamefont {Xue}},\
  }\bibfield  {title} {\enquote {\bibinfo {title} {{Inflation and Alternatives
  with Blue Tensor Spectra}},}\ }\href {\doibase 10.1088/1475-7516/2014/10/075}
  {\bibfield  {journal} {\bibinfo  {journal} {JCAP}\ }\textbf {\bibinfo
  {volume} {10}},\ \bibinfo {pages} {075} (\bibinfo {year} {2014})},\ \Eprint
  {http://arxiv.org/abs/1403.5817} {arXiv:1403.5817 [astro-ph.CO]} \BibitemShut
  {NoStop}%
\bibitem [{\citenamefont {Caprini}\ and\ \citenamefont
  {Sorbo}(2014)}]{Caprini:2014mja}%
  \BibitemOpen
  \bibfield  {author} {\bibinfo {author} {\bibfnamefont {Chiara}\ \bibnamefont
  {Caprini}}\ and\ \bibinfo {author} {\bibfnamefont {Lorenzo}\ \bibnamefont
  {Sorbo}},\ }\bibfield  {title} {\enquote {\bibinfo {title} {{Adding helicity
  to inflationary magnetogenesis}},}\ }\href {\doibase
  10.1088/1475-7516/2014/10/056} {\bibfield  {journal} {\bibinfo  {journal}
  {JCAP}\ }\textbf {\bibinfo {volume} {10}},\ \bibinfo {pages} {056} (\bibinfo
  {year} {2014})},\ \Eprint {http://arxiv.org/abs/1407.2809} {arXiv:1407.2809
  [astro-ph.CO]} \BibitemShut {NoStop}%
\bibitem [{\citenamefont {Mukohyama}\ \emph {et~al.}(2014)\citenamefont
  {Mukohyama}, \citenamefont {Namba}, \citenamefont {Peloso},\ and\
  \citenamefont {Shiu}}]{Mukohyama:2014gba}%
  \BibitemOpen
  \bibfield  {author} {\bibinfo {author} {\bibfnamefont {Shinji}\ \bibnamefont
  {Mukohyama}}, \bibinfo {author} {\bibfnamefont {Ryo}\ \bibnamefont {Namba}},
  \bibinfo {author} {\bibfnamefont {Marco}\ \bibnamefont {Peloso}}, \ and\
  \bibinfo {author} {\bibfnamefont {Gary}\ \bibnamefont {Shiu}},\ }\bibfield
  {title} {\enquote {\bibinfo {title} {{Blue Tensor Spectrum from Particle
  Production during Inflation}},}\ }\href {\doibase
  10.1088/1475-7516/2014/08/036} {\bibfield  {journal} {\bibinfo  {journal}
  {JCAP}\ }\textbf {\bibinfo {volume} {08}},\ \bibinfo {pages} {036} (\bibinfo
  {year} {2014})},\ \Eprint {http://arxiv.org/abs/1405.0346} {arXiv:1405.0346
  [astro-ph.CO]} \BibitemShut {NoStop}%
\bibitem [{\citenamefont {Bastero-Gil}\ \emph {et~al.}(2014)\citenamefont
  {Bastero-Gil}, \citenamefont {Berera}, \citenamefont {Ramos},\ and\
  \citenamefont {Rosa}}]{Bastero-Gil:2014oga}%
  \BibitemOpen
  \bibfield  {author} {\bibinfo {author} {\bibfnamefont {Mar}\ \bibnamefont
  {Bastero-Gil}}, \bibinfo {author} {\bibfnamefont {Arjun}\ \bibnamefont
  {Berera}}, \bibinfo {author} {\bibfnamefont {Rudnei~O.}\ \bibnamefont
  {Ramos}}, \ and\ \bibinfo {author} {\bibfnamefont {Jo\~ao~G.}\ \bibnamefont
  {Rosa}},\ }\bibfield  {title} {\enquote {\bibinfo {title} {{Observational
  implications of mattergenesis during inflation}},}\ }\href {\doibase
  10.1088/1475-7516/2014/10/053} {\bibfield  {journal} {\bibinfo  {journal}
  {JCAP}\ }\textbf {\bibinfo {volume} {10}},\ \bibinfo {pages} {053} (\bibinfo
  {year} {2014})},\ \Eprint {http://arxiv.org/abs/1404.4976} {arXiv:1404.4976
  [astro-ph.CO]} \BibitemShut {NoStop}%
\bibitem [{\citenamefont {Cook}\ and\ \citenamefont
  {Sorbo}(2012)}]{Cook:2011hg}%
  \BibitemOpen
  \bibfield  {author} {\bibinfo {author} {\bibfnamefont {Jessica~L.}\
  \bibnamefont {Cook}}\ and\ \bibinfo {author} {\bibfnamefont {Lorenzo}\
  \bibnamefont {Sorbo}},\ }\bibfield  {title} {\enquote {\bibinfo {title}
  {{Particle production during inflation and gravitational waves detectable by
  ground-based interferometers}},}\ }\href {\doibase
  10.1103/PhysRevD.85.023534} {\bibfield  {journal} {\bibinfo  {journal} {Phys.
  Rev. D}\ }\textbf {\bibinfo {volume} {85}},\ \bibinfo {pages} {023534}
  (\bibinfo {year} {2012})},\ \bibinfo {note} {[Erratum: Phys.Rev.D 86, 069901
  (2012)]},\ \Eprint {http://arxiv.org/abs/1109.0022} {arXiv:1109.0022
  [astro-ph.CO]} \BibitemShut {NoStop}%
\bibitem [{\citenamefont {Carney}\ \emph {et~al.}(2012)\citenamefont {Carney},
  \citenamefont {Fischler}, \citenamefont {Kovetz}, \citenamefont
  {Lorshbough},\ and\ \citenamefont {Paban}}]{Carney:2012pk}%
  \BibitemOpen
  \bibfield  {author} {\bibinfo {author} {\bibfnamefont {Daniel}\ \bibnamefont
  {Carney}}, \bibinfo {author} {\bibfnamefont {Willy}\ \bibnamefont
  {Fischler}}, \bibinfo {author} {\bibfnamefont {Ely~D.}\ \bibnamefont
  {Kovetz}}, \bibinfo {author} {\bibfnamefont {Dustin}\ \bibnamefont
  {Lorshbough}}, \ and\ \bibinfo {author} {\bibfnamefont {Sonia}\ \bibnamefont
  {Paban}},\ }\bibfield  {title} {\enquote {\bibinfo {title} {{Rapid field
  excursions and the inflationary tensor spectrum}},}\ }\href {\doibase
  10.1007/JHEP11(2012)042} {\bibfield  {journal} {\bibinfo  {journal} {JHEP}\
  }\textbf {\bibinfo {volume} {11}},\ \bibinfo {pages} {042} (\bibinfo {year}
  {2012})},\ \Eprint {http://arxiv.org/abs/1209.3848} {arXiv:1209.3848
  [hep-th]} \BibitemShut {NoStop}%
\bibitem [{\citenamefont {Senatore}\ \emph {et~al.}(2014)\citenamefont
  {Senatore}, \citenamefont {Silverstein},\ and\ \citenamefont
  {Zaldarriaga}}]{Senatore:2011sp}%
  \BibitemOpen
  \bibfield  {author} {\bibinfo {author} {\bibfnamefont {Leonardo}\
  \bibnamefont {Senatore}}, \bibinfo {author} {\bibfnamefont {Eva}\
  \bibnamefont {Silverstein}}, \ and\ \bibinfo {author} {\bibfnamefont
  {Matias}\ \bibnamefont {Zaldarriaga}},\ }\bibfield  {title} {\enquote
  {\bibinfo {title} {{New Sources of Gravitational Waves during Inflation}},}\
  }\href {\doibase 10.1088/1475-7516/2014/08/016} {\bibfield  {journal}
  {\bibinfo  {journal} {JCAP}\ }\textbf {\bibinfo {volume} {08}},\ \bibinfo
  {pages} {016} (\bibinfo {year} {2014})},\ \Eprint
  {http://arxiv.org/abs/1109.0542} {arXiv:1109.0542 [hep-th]} \BibitemShut
  {NoStop}%
\bibitem [{\citenamefont {Matarrese}\ \emph {et~al.}(1998)\citenamefont
  {Matarrese}, \citenamefont {Mollerach},\ and\ \citenamefont
  {Bruni}}]{Matarrese:1997ay}%
  \BibitemOpen
  \bibfield  {author} {\bibinfo {author} {\bibfnamefont {Sabino}\ \bibnamefont
  {Matarrese}}, \bibinfo {author} {\bibfnamefont {Silvia}\ \bibnamefont
  {Mollerach}}, \ and\ \bibinfo {author} {\bibfnamefont {Marco}\ \bibnamefont
  {Bruni}},\ }\bibfield  {title} {\enquote {\bibinfo {title} {{Second order
  perturbations of the Einstein-de Sitter universe}},}\ }\href {\doibase
  10.1103/PhysRevD.58.043504} {\bibfield  {journal} {\bibinfo  {journal} {Phys.
  Rev. D}\ }\textbf {\bibinfo {volume} {58}},\ \bibinfo {pages} {043504}
  (\bibinfo {year} {1998})},\ \Eprint {http://arxiv.org/abs/astro-ph/9707278}
  {arXiv:astro-ph/9707278} \BibitemShut {NoStop}%
\bibitem [{\citenamefont {Biagetti}\ \emph {et~al.}(2013)\citenamefont
  {Biagetti}, \citenamefont {Fasiello},\ and\ \citenamefont
  {Riotto}}]{Biagetti:2013kwa}%
  \BibitemOpen
  \bibfield  {author} {\bibinfo {author} {\bibfnamefont {Matteo}\ \bibnamefont
  {Biagetti}}, \bibinfo {author} {\bibfnamefont {Matteo}\ \bibnamefont
  {Fasiello}}, \ and\ \bibinfo {author} {\bibfnamefont {Antonio}\ \bibnamefont
  {Riotto}},\ }\bibfield  {title} {\enquote {\bibinfo {title} {{Enhancing
  Inflationary Tensor Modes through Spectator Fields}},}\ }\href {\doibase
  10.1103/PhysRevD.88.103518} {\bibfield  {journal} {\bibinfo  {journal} {Phys.
  Rev. D}\ }\textbf {\bibinfo {volume} {88}},\ \bibinfo {pages} {103518}
  (\bibinfo {year} {2013})},\ \Eprint {http://arxiv.org/abs/1305.7241}
  {arXiv:1305.7241 [astro-ph.CO]} \BibitemShut {NoStop}%
\bibitem [{\citenamefont {Aghanim}\ \emph {et~al.}(2020)\citenamefont {Aghanim}
  \emph {et~al.}}]{Planck:2018vyg}%
  \BibitemOpen
  \bibfield  {author} {\bibinfo {author} {\bibfnamefont {N.}~\bibnamefont
  {Aghanim}} \emph {et~al.} (\bibinfo {collaboration} {Planck}),\ }\bibfield
  {title} {\enquote {\bibinfo {title} {{Planck 2018 results. VI. Cosmological
  parameters}},}\ }\href {\doibase 10.1051/0004-6361/201833910} {\bibfield
  {journal} {\bibinfo  {journal} {Astron. Astrophys.}\ }\textbf {\bibinfo
  {volume} {641}},\ \bibinfo {pages} {A6} (\bibinfo {year} {2020})},\ \bibinfo
  {note} {[Erratum: Astron.Astrophys. 652, C4 (2021)]},\ \Eprint
  {http://arxiv.org/abs/1807.06209} {arXiv:1807.06209 [astro-ph.CO]}
  \BibitemShut {NoStop}%
\bibitem [{\citenamefont {Kuroyanagi}\ \emph {et~al.}(2015)\citenamefont
  {Kuroyanagi}, \citenamefont {Takahashi},\ and\ \citenamefont
  {Yokoyama}}]{Kuroyanagi:2014nba}%
  \BibitemOpen
  \bibfield  {author} {\bibinfo {author} {\bibfnamefont {Sachiko}\ \bibnamefont
  {Kuroyanagi}}, \bibinfo {author} {\bibfnamefont {Tomo}\ \bibnamefont
  {Takahashi}}, \ and\ \bibinfo {author} {\bibfnamefont {Shuichiro}\
  \bibnamefont {Yokoyama}},\ }\bibfield  {title} {\enquote {\bibinfo {title}
  {{Blue-tilted Tensor Spectrum and Thermal History of the Universe}},}\ }\href
  {\doibase 10.1088/1475-7516/2015/02/003} {\bibfield  {journal} {\bibinfo
  {journal} {JCAP}\ }\textbf {\bibinfo {volume} {02}},\ \bibinfo {pages} {003}
  (\bibinfo {year} {2015})},\ \Eprint {http://arxiv.org/abs/1407.4785}
  {arXiv:1407.4785 [astro-ph.CO]} \BibitemShut {NoStop}%
\bibitem [{\citenamefont {Ben-Dayan}\ \emph {et~al.}(2019)\citenamefont
  {Ben-Dayan}, \citenamefont {Keating}, \citenamefont {Leon},\ and\
  \citenamefont {Wolfson}}]{Ben-Dayan:2019gll}%
  \BibitemOpen
  \bibfield  {author} {\bibinfo {author} {\bibfnamefont {Ido}\ \bibnamefont
  {Ben-Dayan}}, \bibinfo {author} {\bibfnamefont {Brian}\ \bibnamefont
  {Keating}}, \bibinfo {author} {\bibfnamefont {David}\ \bibnamefont {Leon}}, \
  and\ \bibinfo {author} {\bibfnamefont {Ira}\ \bibnamefont {Wolfson}},\
  }\bibfield  {title} {\enquote {\bibinfo {title} {{Constraints on scalar and
  tensor spectra from $N_{eff}$}},}\ }\href {\doibase
  10.1088/1475-7516/2019/06/007} {\bibfield  {journal} {\bibinfo  {journal}
  {JCAP}\ }\textbf {\bibinfo {volume} {06}},\ \bibinfo {pages} {007} (\bibinfo
  {year} {2019})},\ \Eprint {http://arxiv.org/abs/1903.11843} {arXiv:1903.11843
  [astro-ph.CO]} \BibitemShut {NoStop}%
\bibitem [{\citenamefont {Vagnozzi}\ and\ \citenamefont
  {Loeb}(2022)}]{Vagnozzi:2022qmc}%
  \BibitemOpen
  \bibfield  {author} {\bibinfo {author} {\bibfnamefont {Sunny}\ \bibnamefont
  {Vagnozzi}}\ and\ \bibinfo {author} {\bibfnamefont {Abraham}\ \bibnamefont
  {Loeb}},\ }\bibfield  {title} {\enquote {\bibinfo {title} {{The Challenge of
  Ruling Out Inflation via the Primordial Graviton Background}},}\ }\href
  {\doibase 10.3847/2041-8213/ac9b0e} {\bibfield  {journal} {\bibinfo
  {journal} {Astrophys. J. Lett.}\ }\textbf {\bibinfo {volume} {939}},\
  \bibinfo {pages} {L22} (\bibinfo {year} {2022})},\ \Eprint
  {http://arxiv.org/abs/2208.14088} {arXiv:2208.14088 [astro-ph.CO]}
  \BibitemShut {NoStop}%
\bibitem [{\citenamefont {Giar\`e}\ \emph {et~al.}(2023)\citenamefont
  {Giar\`e}, \citenamefont {Forconi}, \citenamefont {Di~Valentino},\ and\
  \citenamefont {Melchiorri}}]{Giare:2022wxq}%
  \BibitemOpen
  \bibfield  {author} {\bibinfo {author} {\bibfnamefont {William}\ \bibnamefont
  {Giar\`e}}, \bibinfo {author} {\bibfnamefont {Matteo}\ \bibnamefont
  {Forconi}}, \bibinfo {author} {\bibfnamefont {Eleonora}\ \bibnamefont
  {Di~Valentino}}, \ and\ \bibinfo {author} {\bibfnamefont {Alessandro}\
  \bibnamefont {Melchiorri}},\ }\bibfield  {title} {\enquote {\bibinfo {title}
  {{Towards a reliable calculation of relic radiation from primordial
  gravitational waves}},}\ }\href {\doibase 10.1093/mnras/stad258} {\bibfield
  {journal} {\bibinfo  {journal} {Mon. Not. Roy. Astron. Soc.}\ }\textbf
  {\bibinfo {volume} {520}},\ \bibinfo {pages} {2} (\bibinfo {year} {2023})},\
  \Eprint {http://arxiv.org/abs/2210.14159} {arXiv:2210.14159 [astro-ph.CO]}
  \BibitemShut {NoStop}%
\bibitem [{\citenamefont {Torrado}\ and\ \citenamefont
  {Lewis}(2021)}]{Torrado:2020dgo}%
  \BibitemOpen
  \bibfield  {author} {\bibinfo {author} {\bibfnamefont {Jesus}\ \bibnamefont
  {Torrado}}\ and\ \bibinfo {author} {\bibfnamefont {Antony}\ \bibnamefont
  {Lewis}},\ }\bibfield  {title} {\enquote {\bibinfo {title} {{Cobaya: Code for
  Bayesian Analysis of hierarchical physical models}},}\ }\href {\doibase
  10.1088/1475-7516/2021/05/057} {\bibfield  {journal} {\bibinfo  {journal}
  {JCAP}\ }\textbf {\bibinfo {volume} {05}},\ \bibinfo {pages} {057} (\bibinfo
  {year} {2021})},\ \Eprint {http://arxiv.org/abs/2005.05290} {arXiv:2005.05290
  [astro-ph.IM]} \BibitemShut {NoStop}%
\bibitem [{\citenamefont {Ade}\ \emph {et~al.}(2021)\citenamefont {Ade} \emph
  {et~al.}}]{BICEP:2021xfz}%
  \BibitemOpen
  \bibfield  {author} {\bibinfo {author} {\bibfnamefont {P.~A.~R.}\
  \bibnamefont {Ade}} \emph {et~al.} (\bibinfo {collaboration} {BICEP, Keck}),\
  }\bibfield  {title} {\enquote {\bibinfo {title} {{Improved Constraints on
  Primordial Gravitational Waves using Planck, WMAP, and BICEP/Keck
  Observations through the 2018 Observing Season}},}\ }\href {\doibase
  10.1103/PhysRevLett.127.151301} {\bibfield  {journal} {\bibinfo  {journal}
  {Phys. Rev. Lett.}\ }\textbf {\bibinfo {volume} {127}},\ \bibinfo {pages}
  {151301} (\bibinfo {year} {2021})},\ \Eprint
  {http://arxiv.org/abs/2110.00483} {arXiv:2110.00483 [astro-ph.CO]}
  \BibitemShut {NoStop}%
\bibitem [{\citenamefont {Smith}\ \emph {et~al.}(2006)\citenamefont {Smith},
  \citenamefont {Pierpaoli},\ and\ \citenamefont
  {Kamionkowski}}]{Smith:2006nka}%
  \BibitemOpen
  \bibfield  {author} {\bibinfo {author} {\bibfnamefont {Tristan~L.}\
  \bibnamefont {Smith}}, \bibinfo {author} {\bibfnamefont {Elena}\ \bibnamefont
  {Pierpaoli}}, \ and\ \bibinfo {author} {\bibfnamefont {Marc}\ \bibnamefont
  {Kamionkowski}},\ }\bibfield  {title} {\enquote {\bibinfo {title} {{A new
  cosmic microwave background constraint to primordial gravitational waves}},}\
  }\href {\doibase 10.1103/PhysRevLett.97.021301} {\bibfield  {journal}
  {\bibinfo  {journal} {Phys. Rev. Lett.}\ }\textbf {\bibinfo {volume} {97}},\
  \bibinfo {pages} {021301} (\bibinfo {year} {2006})},\ \Eprint
  {http://arxiv.org/abs/astro-ph/0603144} {arXiv:astro-ph/0603144} \BibitemShut
  {NoStop}%
\bibitem [{\citenamefont {Cabass}\ \emph {et~al.}(2016)\citenamefont {Cabass},
  \citenamefont {Pagano}, \citenamefont {Salvati}, \citenamefont {Gerbino},
  \citenamefont {Giusarma},\ and\ \citenamefont {Melchiorri}}]{Cabass:2015jwe}%
  \BibitemOpen
  \bibfield  {author} {\bibinfo {author} {\bibfnamefont {Giovanni}\
  \bibnamefont {Cabass}}, \bibinfo {author} {\bibfnamefont {Luca}\ \bibnamefont
  {Pagano}}, \bibinfo {author} {\bibfnamefont {Laura}\ \bibnamefont {Salvati}},
  \bibinfo {author} {\bibfnamefont {Martina}\ \bibnamefont {Gerbino}}, \bibinfo
  {author} {\bibfnamefont {Elena}\ \bibnamefont {Giusarma}}, \ and\ \bibinfo
  {author} {\bibfnamefont {Alessandro}\ \bibnamefont {Melchiorri}},\ }\bibfield
   {title} {\enquote {\bibinfo {title} {{Updated Constraints and Forecasts on
  Primordial Tensor Modes}},}\ }\href {\doibase 10.1103/PhysRevD.93.063508}
  {\bibfield  {journal} {\bibinfo  {journal} {Phys. Rev. D}\ }\textbf {\bibinfo
  {volume} {93}},\ \bibinfo {pages} {063508} (\bibinfo {year} {2016})},\
  \Eprint {http://arxiv.org/abs/1511.05146} {arXiv:1511.05146 [astro-ph.CO]}
  \BibitemShut {NoStop}%
\bibitem [{\citenamefont {Abbott}\ \emph {et~al.}(2021)\citenamefont {Abbott}
  \emph {et~al.}}]{KAGRA:2021kbb}%
  \BibitemOpen
  \bibfield  {author} {\bibinfo {author} {\bibfnamefont {R.}~\bibnamefont
  {Abbott}} \emph {et~al.} (\bibinfo {collaboration} {KAGRA, Virgo, LIGO
  Scientific}),\ }\bibfield  {title} {\enquote {\bibinfo {title} {{Upper limits
  on the isotropic gravitational-wave background from Advanced LIGO and
  Advanced Virgo\textquoteright{}s third observing run}},}\ }\href {\doibase
  10.1103/PhysRevD.104.022004} {\bibfield  {journal} {\bibinfo  {journal}
  {Phys. Rev. D}\ }\textbf {\bibinfo {volume} {104}},\ \bibinfo {pages}
  {022004} (\bibinfo {year} {2021})},\ \Eprint
  {http://arxiv.org/abs/2101.12130} {arXiv:2101.12130 [gr-qc]} \BibitemShut
  {NoStop}%
\bibitem [{\citenamefont {Allen}\ and\ \citenamefont
  {Brustein}(1997)}]{Allen:1996sw}%
  \BibitemOpen
  \bibfield  {author} {\bibinfo {author} {\bibfnamefont {Bruce}\ \bibnamefont
  {Allen}}\ and\ \bibinfo {author} {\bibfnamefont {Ram}\ \bibnamefont
  {Brustein}},\ }\bibfield  {title} {\enquote {\bibinfo {title} {{Detecting
  relic gravitational radiation from string cosmology with LIGO}},}\ }\href
  {\doibase 10.1103/PhysRevD.55.3260} {\bibfield  {journal} {\bibinfo
  {journal} {Phys. Rev. D}\ }\textbf {\bibinfo {volume} {55}},\ \bibinfo
  {pages} {3260--3264} (\bibinfo {year} {1997})},\ \Eprint
  {http://arxiv.org/abs/gr-qc/9609013} {arXiv:gr-qc/9609013} \BibitemShut
  {NoStop}%
\bibitem [{\citenamefont {Wang}\ \emph {et~al.}(2023)\citenamefont {Wang},
  \citenamefont {Zhao}, \citenamefont {Li},\ and\ \citenamefont
  {Zhu}}]{Wang:2023ost}%
  \BibitemOpen
  \bibfield  {author} {\bibinfo {author} {\bibfnamefont {Sai}\ \bibnamefont
  {Wang}}, \bibinfo {author} {\bibfnamefont {Zhi-Chao}\ \bibnamefont {Zhao}},
  \bibinfo {author} {\bibfnamefont {Jun-Peng}\ \bibnamefont {Li}}, \ and\
  \bibinfo {author} {\bibfnamefont {Qing-Hua}\ \bibnamefont {Zhu}},\ }\bibfield
   {title} {\enquote {\bibinfo {title} {{Exploring the Implications of 2023
  Pulsar Timing Array Datasets for Scalar-Induced Gravitational Waves and
  Primordial Black Holes}},}\ }\href@noop {} {\bibfield  {journal} {\bibinfo
  {journal} {.}\ } (\bibinfo {year} {2023})},\ \Eprint
  {http://arxiv.org/abs/2307.00572} {arXiv:2307.00572 [astro-ph.CO]}
  \BibitemShut {NoStop}%
\bibitem [{\citenamefont {Benetti}\ \emph {et~al.}(2022)\citenamefont
  {Benetti}, \citenamefont {Graef},\ and\ \citenamefont
  {Vagnozzi}}]{Benetti:2021uea}%
  \BibitemOpen
  \bibfield  {author} {\bibinfo {author} {\bibfnamefont {Micol}\ \bibnamefont
  {Benetti}}, \bibinfo {author} {\bibfnamefont {Leila~Lobato}\ \bibnamefont
  {Graef}}, \ and\ \bibinfo {author} {\bibfnamefont {Sunny}\ \bibnamefont
  {Vagnozzi}},\ }\bibfield  {title} {\enquote {\bibinfo {title} {{Primordial
  gravitational waves from NANOGrav: A broken power-law approach}},}\ }\href
  {\doibase 10.1103/PhysRevD.105.043520} {\bibfield  {journal} {\bibinfo
  {journal} {Phys. Rev. D}\ }\textbf {\bibinfo {volume} {105}},\ \bibinfo
  {pages} {043520} (\bibinfo {year} {2022})},\ \Eprint
  {http://arxiv.org/abs/2111.04758} {arXiv:2111.04758 [astro-ph.CO]}
  \BibitemShut {NoStop}%
\bibitem [{\citenamefont {Hosseini~Mansoori}\ \emph {et~al.}(2023)\citenamefont
  {Hosseini~Mansoori}, \citenamefont {Felegray}, \citenamefont {Talebian},\
  and\ \citenamefont {Sami}}]{HosseiniMansoori:2023mqh}%
  \BibitemOpen
  \bibfield  {author} {\bibinfo {author} {\bibfnamefont {Seyed~Ali}\
  \bibnamefont {Hosseini~Mansoori}}, \bibinfo {author} {\bibfnamefont
  {Fereshteh}\ \bibnamefont {Felegray}}, \bibinfo {author} {\bibfnamefont
  {Alireza}\ \bibnamefont {Talebian}}, \ and\ \bibinfo {author} {\bibfnamefont
  {Mohammad}\ \bibnamefont {Sami}},\ }\bibfield  {title} {\enquote {\bibinfo
  {title} {{PBHs and GWs from $\mathbb{T}^2$-inflation and NANOGrav 15-year
  data}},}\ }\href@noop {} {\bibfield  {journal} {\bibinfo  {journal} {.}\ }
  (\bibinfo {year} {2023})},\ \Eprint {http://arxiv.org/abs/2307.06757}
  {arXiv:2307.06757 [astro-ph.CO]} \BibitemShut {NoStop}%
\bibitem [{\citenamefont {Cheung}\ \emph {et~al.}(2023)\citenamefont {Cheung},
  \citenamefont {Ouseph},\ and\ \citenamefont {Tseng}}]{Cheung:2023ihl}%
  \BibitemOpen
  \bibfield  {author} {\bibinfo {author} {\bibfnamefont {Kingman}\ \bibnamefont
  {Cheung}}, \bibinfo {author} {\bibfnamefont {C.~J.}\ \bibnamefont {Ouseph}},
  \ and\ \bibinfo {author} {\bibfnamefont {Po-Yan}\ \bibnamefont {Tseng}},\
  }\bibfield  {title} {\enquote {\bibinfo {title} {{NANOGrav Signal and PBH
  from the Modified Higgs Inflation}},}\ }\href@noop {} {\bibfield  {journal}
  {\bibinfo  {journal} {.}\ } (\bibinfo {year} {2023})},\ \Eprint
  {http://arxiv.org/abs/2307.08046} {arXiv:2307.08046 [hep-ph]} \BibitemShut
  {NoStop}%
\bibitem [{\citenamefont {Choudhury}(2023)}]{Choudhury:2023kam}%
  \BibitemOpen
  \bibfield  {author} {\bibinfo {author} {\bibfnamefont {Sayantan}\
  \bibnamefont {Choudhury}},\ }\bibfield  {title} {\enquote {\bibinfo {title}
  {{Single field inflation in the light of NANOGrav 15-year Data:
  Quintessential interpretation of blue tilted tensor spectrum through
  Non-Bunch Davies initial condition}},}\ }\href@noop {} {\bibfield  {journal}
  {\bibinfo  {journal} {.}\ } (\bibinfo {year} {2023})},\ \Eprint
  {http://arxiv.org/abs/2307.03249} {arXiv:2307.03249 [astro-ph.CO]}
  \BibitemShut {NoStop}%
\bibitem [{\citenamefont {Artymowski}\ \emph {et~al.}(2021)\citenamefont
  {Artymowski}, \citenamefont {Ben-Dayan},\ and\ \citenamefont
  {Thattarampilly}}]{Artymowski:2020pci}%
  \BibitemOpen
  \bibfield  {author} {\bibinfo {author} {\bibfnamefont {Micha\l{}}\
  \bibnamefont {Artymowski}}, \bibinfo {author} {\bibfnamefont {Ido}\
  \bibnamefont {Ben-Dayan}}, \ and\ \bibinfo {author} {\bibfnamefont
  {Udaykrishna}\ \bibnamefont {Thattarampilly}},\ }\bibfield  {title} {\enquote
  {\bibinfo {title} {{Sourced fluctuations in generic slow contraction}},}\
  }\href {\doibase 10.1088/1475-7516/2021/06/010} {\bibfield  {journal}
  {\bibinfo  {journal} {JCAP}\ }\textbf {\bibinfo {volume} {06}},\ \bibinfo
  {pages} {010} (\bibinfo {year} {2021})},\ \Eprint
  {http://arxiv.org/abs/2011.00626} {arXiv:2011.00626 [gr-qc]} \BibitemShut
  {NoStop}%
\bibitem [{\citenamefont {Ben-Dayan}\ and\ \citenamefont
  {Thattarampilly}(2023)}]{Ben-Dayan:2023rlj}%
  \BibitemOpen
  \bibfield  {author} {\bibinfo {author} {\bibfnamefont {Ido}\ \bibnamefont
  {Ben-Dayan}}\ and\ \bibinfo {author} {\bibfnamefont {Udaykrishna}\
  \bibnamefont {Thattarampilly}},\ }\bibfield  {title} {\enquote {\bibinfo
  {title} {{Requiem to ''Proof of Inflation'' or Sourced Fluctuations in a
  Non-Singular Bounce}},}\ }\href@noop {} {\bibfield  {journal} {\bibinfo
  {journal} {.}\ } (\bibinfo {year} {2023})},\ \Eprint
  {http://arxiv.org/abs/2308.00256} {arXiv:2308.00256 [astro-ph.CO]}
  \BibitemShut {NoStop}%
\end{thebibliography}%
\end{document}